\documentclass[%
reprint,
noeprint,
superscriptaddress,
amsmath,amssymb,
prl,
]{revtex4-2}

\usepackage{mathtools}	
\usepackage{times}
\usepackage[colorlinks=true,linkcolor=blue,urlcolor=blue,citecolor=blue]{hyperref}

\usepackage{catchfilebetweentags}
\newif\ifsupp
\supptrue

\usepackage{graphicx}% Include figure files
\usepackage{dcolumn}% Align table columns on decimal point
\usepackage{bm}% bold math
\usepackage{float}
\usepackage{braket}
\usepackage{xcolor}

\newcommand{\abs}[1]{\lvert#1\rvert}

\newcommand{\nn}{\nonumber}

\begin{document}
	\title{Resonant enhancement of three-body loss between  strongly interacting photons} 
	
	\author{Marcin Kalinowski}
		\thanks{These two authors contributed equally.}
	\affiliation{Joint Quantum Institute, NIST/University of Maryland, College Park, Maryland 20742 USA}
	\affiliation{Faculty of Physics, University of Warsaw, Pasteura 5, 02-093 Warsaw, Poland}
	\author{Yidan Wang}
	\thanks{These two authors contributed equally.}
	\affiliation{Joint Quantum Institute, NIST/University of Maryland, College Park, Maryland 20742 USA}
	 \author{Przemyslaw Bienias}
	 \affiliation{Joint Quantum Institute, NIST/University of Maryland, College Park, Maryland 20742 USA}
	 \affiliation{Joint Center for Quantum Information and Computer Science, NIST/University of Maryland, College Park, Maryland 20742 USA}
	 \author{ Michael J. Gullans}
	 \affiliation{Joint Quantum Institute, NIST/University of Maryland, College Park, Maryland 20742 USA}
	 \affiliation{Joint Center for Quantum Information and Computer Science, NIST/University of Maryland, College Park, Maryland 20742 USA}
	 \affiliation{Department of Physics, Princeton University, Princeton, New Jersey 08544 USA}
	 \author{Dalia P. Ornelas-Huerta}	 
	 \affiliation{Joint Quantum Institute, NIST/University of Maryland, College Park, Maryland 20742 USA}
	 \author{ Alexander N. Craddock}
	 \affiliation{Joint Quantum Institute, NIST/University of Maryland, College Park, Maryland 20742 USA}
	 \author{Steven L. Rolston}
	  \affiliation{Joint Quantum Institute, NIST/University of Maryland, College Park, Maryland 20742 USA}
	 \author{J. V. Porto}
	 \affiliation{Joint Quantum Institute, NIST/University of Maryland, College Park, Maryland 20742 USA}
	 \author{ Hans Peter B{\" u}chler}
	 \affiliation{Institute for Theoretical Physics III and Center for Integrated Quantum Science and Technology, University of Stuttgart, 70550 Stuttgart, Germany}
	 \author{Alexey V. Gorshkov}
	 \affiliation{Joint Quantum Institute, NIST/University of Maryland, College Park, Maryland 20742 USA}
	 \affiliation{Joint Center for Quantum Information and Computer Science, NIST/University of Maryland, College Park, Maryland 20742 USA}

	\date{\today}
	
	\begin{abstract}
		Rydberg polaritons provide an example of a rare type of system where three-body interactions can be as strong or even stronger than two-body interactions. The three-body interactions can be either dispersive or dissipative, with both types possibly giving rise to exotic, strongly-interacting, and topological phases of matter. Despite past theoretical and experimental studies of the regime with dispersive interaction,  the dissipative regime is still mostly unexplored. Using a  renormalization group technique to solve the three-body Schr{\" o}dinger equation, we show how the shape and strength of dissipative three-body forces can be universally enhanced for Rydberg polaritons.  We demonstrate how these interactions relate to the transmission through a single-mode cavity, which can be used as a probe of the three-body physics in current experiments.
	\end{abstract}

	\maketitle
	{\it Introduction.---}Systems exhibiting strong interactions between single photons are an exciting frontier of quantum optics \cite{Chang2014}.  They are practically relevant for quantum networks \cite{Kimble2008} and can give rise to new exotic states of matter \cite{Carusotto2013,Maghrebi2015,Otterbach2013}. Obtaining better control and understanding  of these systems in the quantum few-body limit is central to realizing this potential in near-term experiments.
	An important step in this direction is the mastery of the three-body problem. Although in general not analytically solvable, the three-body problem has emergent universal properties, such as the existence of Efimov bound states~\cite{Efimov1970}. 
	Moreover, three-body forces can greatly influence the properties of quantum many-body systems as in the case of nuclear systems~\cite{Brown1969}, neutron stars~\cite{Steiner2012}, and fractional quantum Hall states~\cite{Moore1991}.
	
	By coupling photons to Rydberg atoms using electromagnetically induced transparency (EIT) \cite{Fleischhauer2005}, strong and tunable pairwise interactions between photons are achievable~\cite{Dudin2012, Peyronel2012, Maxwell2013,Stanojevic2013,Li2016,Gorniaczyk2014, Tiarks2014, Gorniaczyk2016,Tiarks2016, Thompson2017,Tiarks2019,Firstenberg2013,Liang2018,Stiesdal2018,Sommer2016,Schine2016,Jia2018,Clark2019b,Cantu2020}. 
	Recently, it has been demonstrated that three-body forces between Rydberg polaritons can be very strong as well~\cite{Jachymski2016, Gullans2016,Gullans2017,Liang2018,Bienias2020a}, which distinguishes them from weaker three-body forces engineered with ultracold atoms~\cite{Gambetta2020} and molecules~\cite{Buchler2007a,Daley2009,Johnson2009,Mazza2010}.
	However, the study of dissipative three-body interactions has only begun to be explored \cite{Ornelas-Huerta2020b}. 
	Dissipative forces are of interest as they often lead to exotic nonequilibrium dynamics in driven-dissipative systems~\cite{Peyronel2012,Gorshkov2013,Tresp2015,Zeuthen2017,Bienias2020c}, 
	while also finding applications in engineering topological phases of matter such as the Pfaffian state \cite{Roncaglia2010}.
	
\begin{figure}[t]
		\centering
		\includegraphics[width=.48\textwidth]{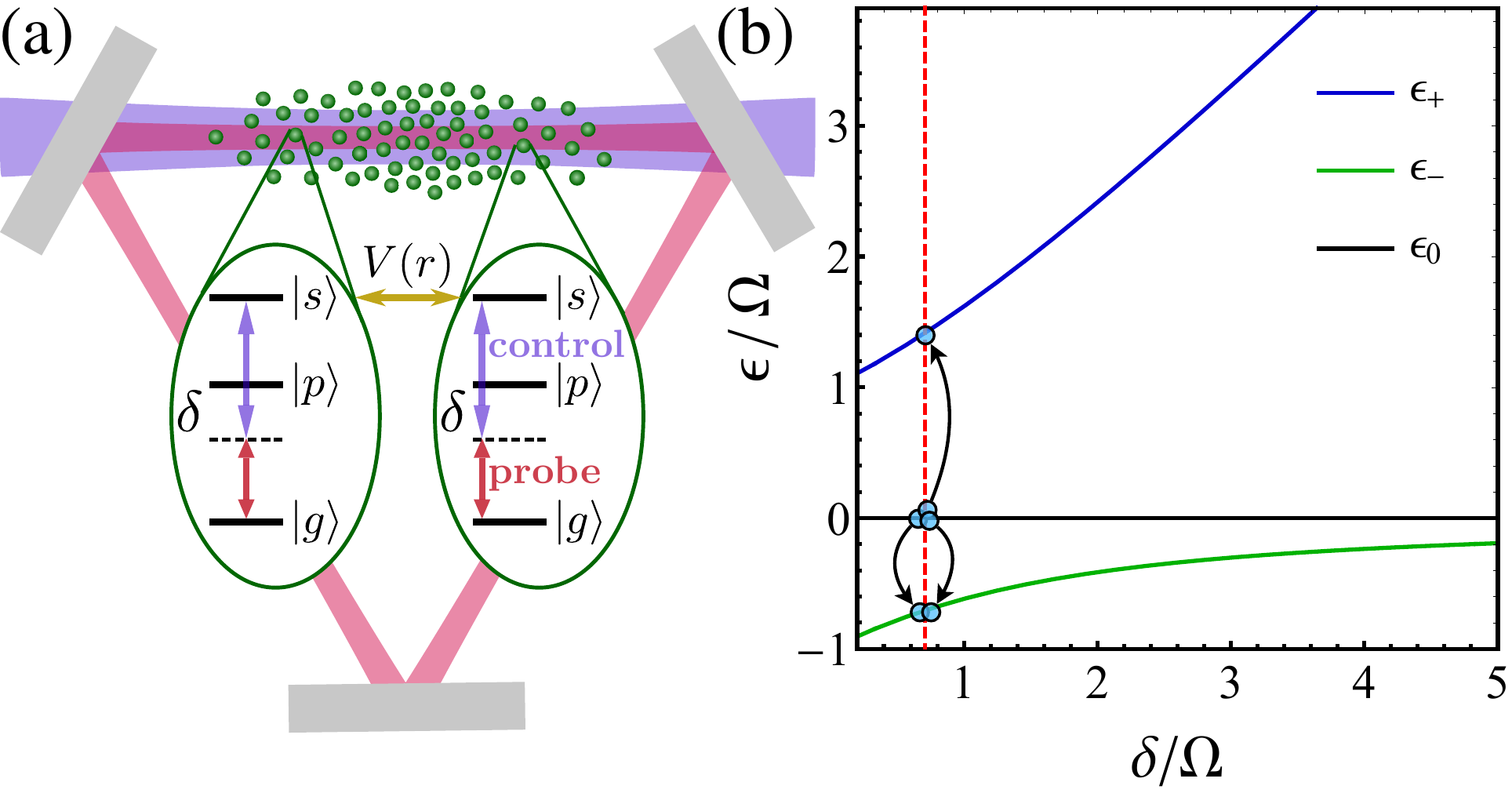}
		\caption{(a) Gas of neutral atoms is suspended in an optical cavity. Each atom is a three-level system with the ground state $\ket{g}$, intermediate lossy state $\ket{p}$ with half-width $\gamma$, and a high-lying Rydberg state $\ket{s}$, which experiences strong interactions. Classical control field with Rabi frequency $\Omega$ and detuning $\delta$ couples states $\ket{p}$ and $\ket{s}$. Quantum photon field 	with collective coupling $g$ drives the  $\ket{g}-\ket{p}$ transition and is tuned to the two-photon resonance.  (b) Energy of the upper (blue) and lower (green) branches of spin waves as a function of the single-photon detuning. At $\delta = \Omega/\sqrt{2}$ scattering of three DSPs into spin waves is on resonance. }
		\label{fig:sys}
	\end{figure}

	In this Letter, we study the influence of dissipative three-body interactions on the physics of Rydberg cavity polaritons.  
	Pure three-body scattering processes in Rydberg-EIT systems are strong and often comparable to two-body effects \cite{Jachymski2016,Gullans2016,Liang2018,Stiesdal2018,Ornelas-Huerta2020b}.  There is also evidence that effective three-body interactions are enhanced in this system \cite{Jachymski2016,Gullans2016,Liang2018} due to Rydberg blockade effects \cite{Lukin01}. 
	Here, by studying a simplified cavity model  that can be treated with a rigorous renormalization group technique, we clearly establish  the existence of a universal regime where both dispersive and dissipative three-body forces can be  enhanced in a tunable fashion.  This enhancement appears due to a near-resonant process when the incoming state can conserve energy and momentum by scattering to a large manifold of intermediate lossy states [see Fig.~\ref{fig:sys}(b)].  Due to the role played by an intermediate resonant channel, this effect has  similarities to Feshbach resonances~\cite{Lewenstein2007}.    The interaction can be tuned using the strength and the frequency of the classical control fields.  We show how these effects can be probed in current experiments by studying the cavity transmission.
	
	Because of the multi-component nature of the Rydberg polaritons, the theoretical description of the three-body problem is nuanced and complex.   To make our analysis analytically tractable, we concentrate on a single-mode cavity, with the extensions to multi-mode cavities presented in~our upcoming work \cite{Kalinowski2020}.  Specifically, based on the microscopic model of photons in a    cavity~\cite{Sommer2015,Parigi2012,Stanojevic2013,Georgakopoulos2018,Litinskaya2016,Grankin2014}
	interacting with Rydberg atoms under EIT conditions, we derive analytical formulae for the interaction-induced shifts in energies and decay rates of three dark-state polaritons (DSPs).  We show how to gain additional insights into the system by introducing an effective Hamiltonian describing dark state polaritons alone--this approach may be useful in deriving effective descriptions of the free-space system.  We solve the three-body problem  using a simplified version of the Faddeev equations. 
The methods introduced here allow one to extend the analysis of Ref.~\cite{Jachymski2016} to compute energy shifts for arbitrary multi-mode cavities and improve the accuracy of the extracted three-body force. They  may also aid in developing a more systematic and rigorous renormalization group approach for the free-space problem \cite{Gullans2016}.
	
	{\it System.---}The medium we consider consists of three-level atoms with ground-state $\ket{g}$, and an intermediate state $\ket{p}$ coupled to Rydberg state $ \ket{s} $ by a coherent laser, with Rabi frequency $\Omega$, and a complex detuning $\Delta=\delta-i\gamma$ [see Fig.~\ref{fig:sys}(a)], which captures the $\ket{p}$-state's decay rate $2\gamma$. The atomic cloud is suspended in a single-mode running-wave cavity.  The quantum photon field, with collective coupling $g$, is tuned to the EIT resonance with the noninteracting Hamiltonian
 \begin{align} \label{eqn:Hcav}
H_0 & =  \int dz\,\vec{\psi}^\dag(z) \left(
\begin{array}{c c c}
0&g & 0 \\
g & \Delta  & \Omega \\
0 & \Omega & 0
\end{array} \right) \vec{\psi}(z) ,
 \end{align}
where $\vec{\psi}^\dag(z)= [u_0^\ast(z) a^\dag,P^\dag(z),S^\dag(z)]$ is a vector of bosonic creation operators for the cavity field $a$ with mode function $u_0(z)$ and atomic states $\ket{p},\ket{s}$ at position $z$. We set $\hbar=1$ throughout. $H_0$ couples the cavity field to one $\ket{p}$ mode and one $\ket{s}$ mode, both with the same mode function $u_0(z)$. Diagonalizing the resulting 3-by-3 matrix leads to three eigenmodes.
The zero-energy mode is the DSP, which has no overlap with the lossy intermediate state.  The two ``bright-state" polariton modes are energetically separated and do not influence the DSP behavior in the experimentally relevant limit of strong coupling ($ g\rightarrow\infty $) considered here. The remaining eigenstates of $H_0$ (spin waves) correspond to the excitations of the atomic cloud, have no photonic component, and couple to the polaritons only via atom-atom interactions.  In the presence of Rydberg interactions $H_{\rm int} = \frac{1}{2}\int dz dz' S^\dag(z)S^\dag(z') V(z-z') S(z')S(z)$, polaritons experience an effective two-body potential \cite{Bienias2014,Gorshkov2011,Bienias2020}
	\begin{equation}
	\label{eq:Veff2}
	U_{2}(\omega;r) = \frac{V(r)}{1-\chi(\omega)V(r)},
	\end{equation}
	where $\chi(\omega)$ is a function of $\Delta,\Omega$, and the total energy $\omega$ of the incoming polaritons. The bare interaction ${V(r)=C_6/r^6}$ is the van der Waals  potential between two atoms separated by distance $r$. The effective potential in Eq.~\eqref{eq:Veff2} saturates to a constant value at short distances and decays as $1/r^6$ for large separations. Intuitively, at large distances, the van der Waals interaction is directly transferred onto the polaritons, while at distances shorter than the blockade radius $ r_b=\abs{\chi(0)C_6}^{1/6} $, the interaction shifts the two Rydberg states out of resonance (the so-called Rydberg blockade mechanism) leading to the renormalization of the  effective interaction potential.  Previous works on the three-body problem considered the restricted limit $\Omega \ll |\delta|$ \cite{Jachymski2016,Gullans2016,Gullans2017}; here, we extend the regime of applicability to $\Omega>|\delta|$, which allows for a more general description of the system, including repulsive photons \cite{Bienias2014} and dissipative behavior.

	Three-body problems are complex both in classical and quantum physics. To gain insight into few-body interactions, we consider a cavity as our setup, since its treatment requires only a finite number of photonic modes.  When the photonic modes are near-degenerate (or there is only one relevant photonic mode), there is a natural separation of scale that appears between low-energy polaritons and high-energy atomic excitations (spin waves). 
	We can take advantage of this energy separation to obtain an effective theory for the polaritons--renormalized by the influence of high-energy spin waves. In contrast, in free space, there is a continuum of energies connecting these two regimes, which makes a similar procedure  more difficult.
	 
	For simplicity, we consider an effectively one-dimensional running-wave cavity with a single, fixed-momentum photonic mode on EIT resonance and a uniform density of atoms filling the entire cavity mode.  We present the generalization of our results to nonuniform setups, e.g., as in Fig.~\ref{fig:sys}(a), in our upcoming work \cite{Kalinowski2020}. We focus on this model because we have found that it captures generic physical features of multi-mode systems, while  simplifying certain technical aspects of the calculations.
	
	Independent of the geometry, in such cavity models, the interactions between polaritons  most simply appear as shifts in the energies and decay rates of the polariton modes. To calculate these shifts, we use a master equation description of the problem in the weak-driving regime, such that the anti-Hermitian part of the non-Hermitian Hamiltonian is sufficient to account for losses in the system \cite{Carmichael1993}.   At a technical level, complex energy shifts for the two- and three-body problems coincide with the value of poles in the corresponding two- and three-body $T$-matrix describing correlation functions in this  system. Here, we present only the important steps of the derivation, while detailed calculations are included in the Supplemental Material \cite{supplement}.
	
\begin{figure}[t]
		\centering
		\includegraphics[width=\linewidth]{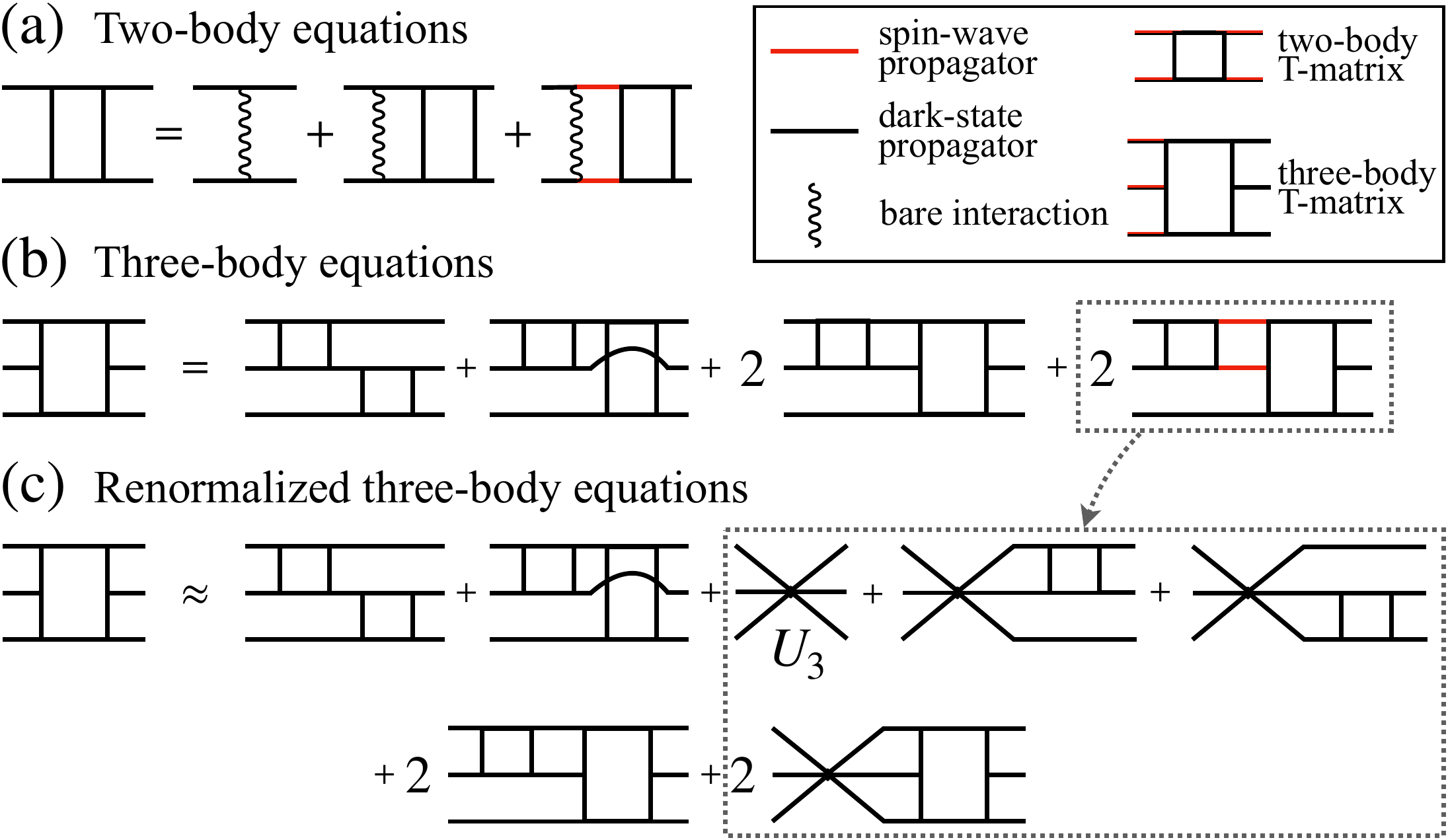}
		\caption{(a) Diagrammatic representation (see inset for legend) of the Lippmann-Schwinger equation for the scattering of two dark-state polaritons. Processes with only one spin wave are forbidden by momentum conservation. (b) Schematic representation of Faddeev equations for the three-body $T$-matrix $ T_3^{12}(\omega) $, where particles 1 and 2 interact first. (c) Truncation of the three-body equations (b)  to second order in $r_b/L$ allows for expressing the spin-wave contribution in the form of an effective three-body potential $ U_3 $ between DSPs.}
		\label{fig:T2T3diag}
	\end{figure}

	{\it Two-body problem.---}First, we turn to the two-body problem as the result is required as input to our solution for the three-body problem.  
	Consider an incoming state of two polaritons (labeled 1 and 2) initially located at positions $\vec{x}=(x_1,x_2)$
	and later measured at positions $\vec{x}'$ after interactions take place.    The amplitude for this process can be described within the framework of scattering theory.
	The multi-component nature of the polariton problem means that the full (bold) two-body $T$-matrix $ \bm{\hat{T}_2}(\omega) $ is a $ 3\times3 $ operator-valued matrix. However, only the Rydberg ($ \ket{s}$)  component experiences interactions.
	Therefore, we can restrict our considerations to the $ ss $-component $ \hat{T}_2(\omega) $ of the full two-body $T$-matrix $\bm{\hat{T}_2}(\omega)$ \cite{Bienias2014,Bienias2016,Bienias2016a}. 
	$ \hat{T}_2(\omega) $ is governed by the Lippmann-Schwinger equation, see Fig.~\ref{fig:T2T3diag}(a). {In the running-wave cavity with a uniform atomic density, the problem is simplified because both the polariton and spin-wave modes become plane waves. In addition, due to the general fact that propagators do not depend on momentum in these cavity problems, one can equivalently study the integrated $T$-matrix $T_2(\omega)\equiv\int d\vec{x}d\vec{x}'\, T_2(\omega;\vec{x},\vec{x}')$.} Then, the solution of the Lippmann-Schwinger equation is  
	\begin{equation}\label{eq:T2eq}     
	T_2(\omega) = \frac{\omega\,U_2(\omega)}{\omega-U_2(\omega)[1-\omega\,\chi(\omega)]},
	\end{equation}
	where $U_2\equiv\int \frac{dr}{L}U_2(\omega;r)$ and $L$ is the  mode volume of the DSP.  One can obtain an exact equation for the two-body energy shift by solving for the poles of $T_2(\omega)$.
	
	{\it Three-body problem.---}Although dramatically simplified, the three-body problem cannot, to the best of our knowledge, be solved exactly in this single-mode cavity model.  Instead, we approximate the full result by a power series in the small parameter $r_b/L$, which is effectively the 
	product of the blockade radius and the density of polaritons in this few-body limit.  In experimental realizations, this condition is often satisfied, which is an additional motivation for working within this expansion. We stress, that this approach is still nonperturbative in the bare interaction $ V(r) $. The three-body energy shift to second order in $r_b/L$ is given by
	\begin{equation}\label{eq:E3series}
	\delta E_3 = \frac{ r_b}{L}E_3^{(1)}+ \frac{r_b^2}{L^2}E_3^{(2)}+\mathcal{O}(r_b^3/L^3),
	\end{equation}
	and our goal is to derive coefficients $ E_3^{(1/2)}$.
	Similarly to the two-body case, we consider just the $sss$-component $\hat{T}_3$ of the full three-body $T$-matrix $\bm{\hat{T}_3}$ (bold). 
	
	The three-body quantum scattering problem can be recast as an infinite series of two-body interactions using Faddeev equations \cite{Faddeev1961}. In this formalism, all scattering processes are grouped depending on which pair of particles interacts first. Crucially, the $T$-matrix separates into the sum
	\begin{equation}
	\hat{T}_3(\omega) = \frac{1}{3}\sum_{i<j} \hat{T}_3^{ij}(\omega,\epsilon_k),
	\end{equation} 
	where $\hat{T}_3^{ij}(\omega,\epsilon_k)$ denotes the $T$-matrix for scattering where particles $i,j$ interact first and the third particle $ k\neq i,j $ has incoming energy $ \epsilon_k $. The equation for $\hat{T}_3^{12}(\omega,\epsilon)$, when all outgoing states are DSPs, is
	\begin{align}\label{eq:Fadeev}
	&\hat{T}_{3}^{12}(\omega,\epsilon) =   \hat{T}^{12}_2(\omega-\epsilon)\hat{g}_s(\omega)[\hat{T}^{23}_2(\omega)+\hat{T}^{13}_2(\omega)]+\\&\int d\tilde{\epsilon}\,\hat{T}_2^{12}(\omega-\epsilon)\hat{g}_s(\tilde{\epsilon})\hat{g}_s(\omega-\epsilon-\tilde{\epsilon}) [\hat{T}^{23}_3(\omega,\tilde{\epsilon})+\hat{T}^{13}_3(\omega,\tilde{\epsilon})],\nn
	\end{align}
	where $ \hat{T}_2^{13} $ describes the two-body scattering of particles labeled $ 1 $, $ 3 $. 
	Equations for the other $\hat{T}_3^{ij}$ can be obtained by permutation of particles. The Rydberg-component propagator $ \hat{g}_s $ is a complex object that involves contributions from different spin-wave branches and the DSP mode. It is also restricted by momentum conservation, which forbids single spin-wave excitations. Note that the simple form of Eq.~\eqref{eq:Fadeev} is thanks to the use of abstract operators. The representation in e.g. a coordinate basis is more involved \cite{supplement}. 
	
	To derive an effective DSP theory, we separate spin-wave and DSP components in $ \hat{g}_s $, which will allow us to perform expansion in $ r_b/L $. The equation for the $T$-matrix describing DSP-to-DSP scattering $ \hat{T}_3(\omega)\equiv\hat{T}_3^{12}(\omega,0) $ is represented diagrammatically in Fig.~\ref{fig:T2T3diag}(b), where we explicitly showed separated spin-wave (red) and DSP (black) propagators. Next, we restrict both sides of Eq.~\eqref{eq:Fadeev} to the second order in $r_b/L$.  To understand which contributions can be neglected, we examine constituents of our diagrams. 
	{First, each two-body vertex gives a factor of $r_b/L$, which follows from Eq.~\eqref{eq:T2eq} because ${T_2 \sim r_b/L}$. Second, whenever the summation over a macroscopic number of spin waves is present, the propagator introduces a factor of $L$. 
	Third, when all three intermediate excitations are DSPs the propagator equals $[\omega+i0^+]^{-1}$, which is of  order $L/{r_b}$, since the energy shift satisfies $\omega\sim r_b/L$. All other contributions are $ \mathcal{O}(1) $. 
	Finally, we rewrite the original Faddeev equations in an approximate form shown in Fig.~\ref{fig:T2T3diag}(c) -- without any spin-wave degrees of freedom.} The leading-order contribution due to spin waves is contained in the effective three-body potential $U_{3}(\omega)$, which is represented by a star-like diagram in Fig.~\ref{fig:T2T3diag}(c).  The equations defining $U_{3}(\omega) = \int d^3 x\, U_{3}^0(\omega;\vec{x})$ are expressed in terms of matrix equations for the integrand $U_{3}^0(\omega;\vec{x})$, which can be solved analytically for each $\vec{x}$ and then integrated numerically to get the value of $ U_3(\omega) $.  We provide the full set of equations in the Supplemental Material \cite{supplement}.  Similarly to the two-body case, we obtain the equation for the energy shift in the system as a pole in $ T_3(\omega) $, which we solve self-consistently to obtain the expansion of $ \delta E_3 $ as in Eq.~\eqref{eq:E3series}.
	
	From the analytic structure of our solution for $T_3$ and $U_{3}$, we predict a strong enhancement ($\sim \! \Omega/\gamma$) of three-body losses in the vicinity of the resonance condition $\delta = \Omega/\sqrt{2}$.  This enhancement phenomenon has an intuitive explanation:  three dark-state polaritons propagating at EIT resonance have zero energy. At the same time, upper and lower branches of spin waves have energies $\epsilon_{\pm}= \frac{1}{2}(\Delta \pm \sqrt{\Delta^2+4 \Omega^2})$. Strong losses occur at $\delta = \Omega/\sqrt{2}$ because, at this point, ${2\epsilon_-+\epsilon_+=0}$, which means that three-body scattering into lossy atomic excitations is on energetic resonance [see Fig.\ref{fig:sys}(b)].
	
	\begin{figure}[bt]
		\centering
		\includegraphics[width=\linewidth]{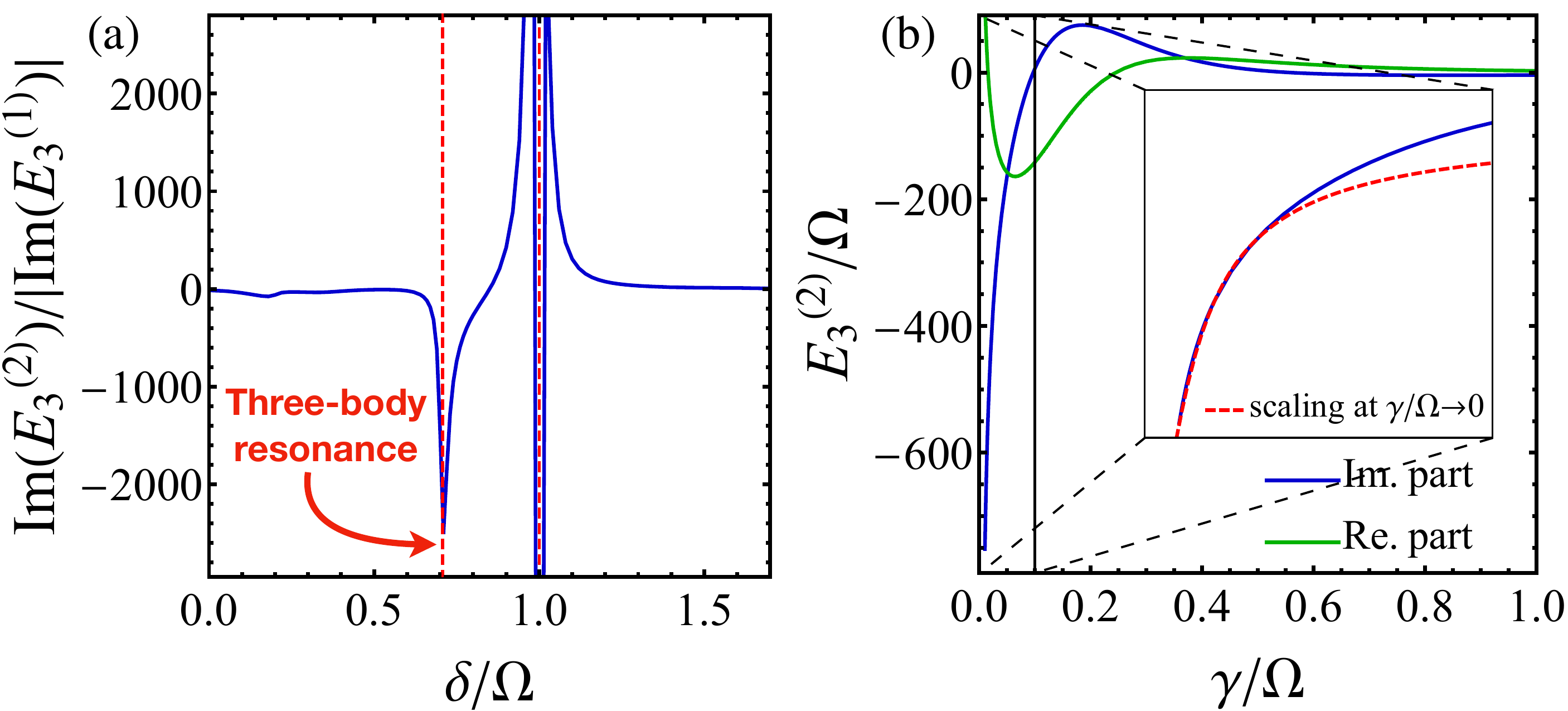}
		\caption{(a) Ratio of imaginary parts of $E_{3}^{(2)}$ and $E_{3}^{(1)}$
		at $\gamma/\Omega=0.01$. Near $\delta/\Omega=1/\sqrt{2}$, we observe enhancement caused by a three-body resonance.  $\delta = \Omega$ is a singular point where $r_b \to 0$.   (b) Real and imaginary parts of $E_3^{(2)}$ 
		as a function of $\gamma/\Omega$ at $\delta/\Omega=1/\sqrt{2}$. (inset) Our numerical results (blue) agree with analytical scaling  arguments (dashed red) suggesting that the ratio $\textrm{Im}(E_3^{(2)})/\textrm{Re}(E_3^{(2)})$ diverges as $\Omega/\gamma$ in the limit $\gamma/\Omega \rightarrow 0$.}
		\label{fig:E3E2num}
	\end{figure}

	In Fig.~\ref{fig:E3E2num}(a), we characterize the strength of three-body loss using the ratio of the expansion coefficients $\textrm{Im}(E_3^{(2)})/\textrm{Im}(E_3^{(1)})$. The denominator $\textrm{Im}(E_3^{(1)})$ from Eq.~\eqref{eq:E3series}  contains contributions to three-body loss from disconnected two-body processes only. 
	We see the predicted enhancement at the resonance condition $\delta = \Omega/\sqrt{2}$.  There is an additional resonant feature at $\delta = \Omega$ that arises because of a two-body interference effect whereby $\chi$ vanishes and, therefore, $r_b$ goes to zero.  This leads to an overall enhancement of both two and three-body interaction effects, which makes this regime difficult to analyze.  In contrast, for the three-body resonance condition $\delta = \Omega/\sqrt{2}$, there are no resonant features that appear in the two-body problem.  Therefore, we interpret the enhancement of the three-body loss observed at this point as a genuinely three-body effect.  
	
	In Fig.~\ref{fig:E3E2num}(b), we show the dependence of $E_3^{(2)}$ on the decay rate $\gamma$ at the resonance condition $\delta = \Omega/\sqrt{2}$.  We find a divergence as $\gamma/
	\Omega \rightarrow 0$ (see inset), consistent with above analytic scaling arguments, indicating that the enhancement factor for three-body loss can be made arbitrarily large.  We note, however, that our calculations only apply when the perturbative expansion in Eq.~\eqref{eq:E3series} is valid.   As a result, we cannot definitively say whether the three-body loss dominates over two-body loss for finite $r_b/L$ because we have not obtained any estimates or bounds on the higher order terms in the expansion.  
	
	As we discuss in the follow-up work \cite{Kalinowski2020}, this enhancement is independent of the details of the photonic mode geometry; therefore, we interpret it as a universal effect for Rydberg polariton systems. Because this regime of enhanced losses is  prevalent across different cavity setups, it is a promising candidate to realize physical systems driven by  three-body interactions.  The significance of this three-body resonance to free-space transmission is difficult to determine due to the increased complexity of the free-space problem. Recently, we experimentally studied an enhanced three-body loss feature in free-space that occurs in a similar parameter regime $\Omega \sim \delta$, but has  a richer physical origin \cite{Ornelas-Huerta2020b}.
	
	\begin{figure}
		\centering
		\includegraphics[width=\linewidth]{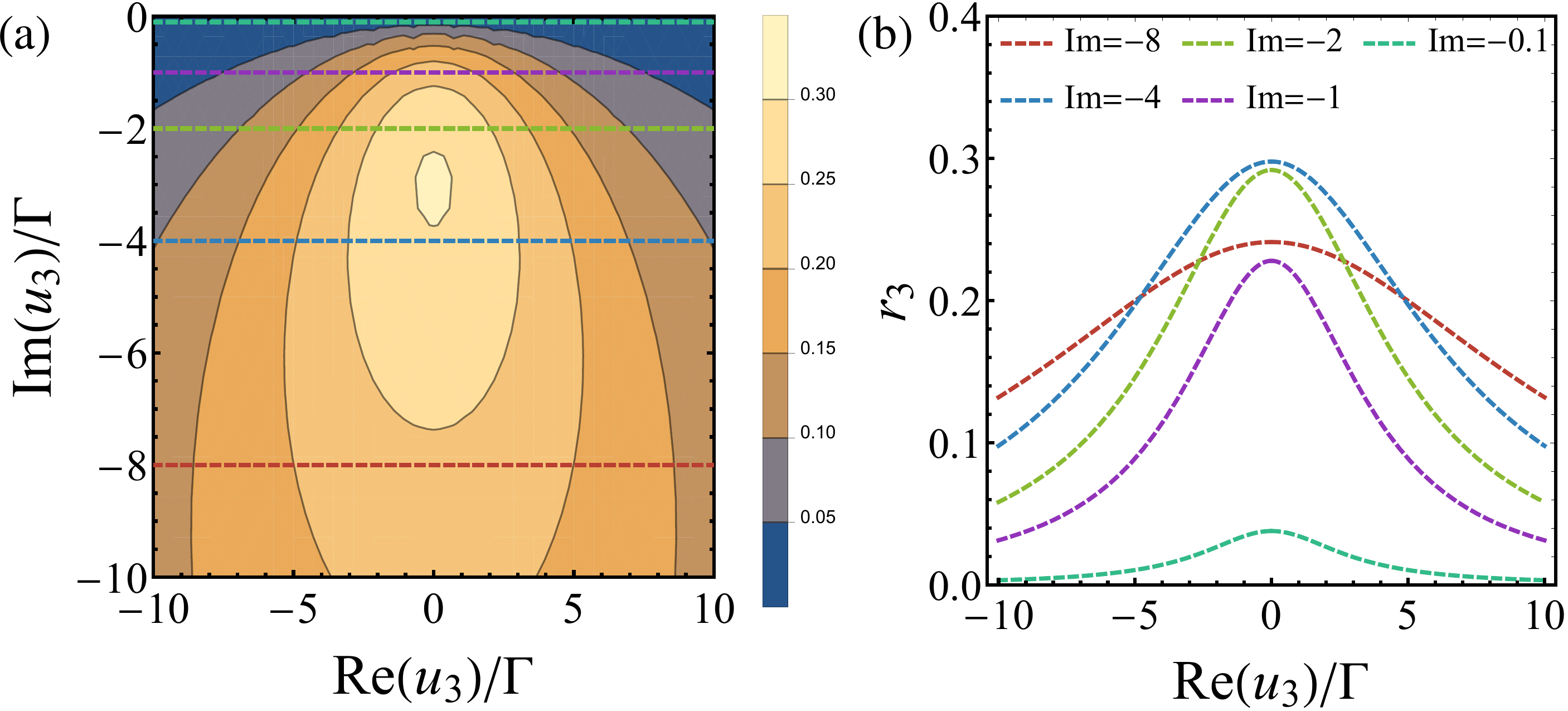}
		\caption{(a-b) Three-body loss parameter $r_3$ in units of $\Gamma^{-2}$ as a function of real and imaginary part of $u_3/\Gamma$ for the case of $u_2=0$, which corresponds to negligible two-body correlations in the microscopic model. A local maximum appears in $r_3$ due to a blockade of the cavity when three-body interactions are strong. The five curves in (b) correspond to the five values of $\textrm{Im}(u_3)$ indicated by the horizontal lines in (a).}
		\label{fig:transmission}
	\end{figure}{}
	
	{\it Experimental probing.---}In order to relate our microscopic description to experimentally measurable quantities, we now study the transmission through our cavity system.  We use a low-energy model for the transmission where the only excitations in the cavity coupled to the waveguide are the DSPs. The effective Hamiltonian for the cavity-DSPs is
	\begin{equation}\label{eq:Htr}
	H = -i (\Gamma+\kappa) b^\dagger b + u_2 (b^\dagger)^2 b^2+u_3 (b^\dagger)^3b^3,
	\end{equation}
	where $b^\dag$ is a bosonic creation operator for the DSPs, $2\Gamma$ is the decay rate of DSPs from the cavity into the waveguide, $2\kappa$ is the decay rate to other modes, and the coefficients $u_2,u_3$ are related to energy shifts $\delta E_2,~\delta E_3$, calculated as described above from the full microscopic theory, through $ {u_2 = \delta E_2}$ and ${u_3=\delta E_3 - 3\,\delta E_2}$.  For  simplicity, we focus on the limit where three-body effects dominate over two-body phenomena by taking $u_2=0$ in Eq.~\eqref{eq:Htr}. We use a measure of three-body loss that is appropriate when all decay is into the  waveguide ($ \kappa = 0   $) and when two-body interactions $u_2$ are zero or negligibly small:
	\begin{equation}\label{eq:r3}
	r_3 = \int d\tau_1 d\tau_2 \,[1 - g^{(3)}(\tau_1,\tau_2) ],   
	\end{equation}
	where $g^{(3)}(\tau_1,\tau_2)$ is the three-photon correlation function at the output of the waveguide for a weak, continuous-wave coherent-state input and where $\tau_{1,2}$ are the relative coordinates of the three photons.  The $r_3$ parameter measures the probability that three-photons are lost from the pulse due to the interactions.
	  We can analytically compute $r_3$  for this transmission problem as detailed in the Supplemental Material \cite{supplement}.  We have also extended these calculations to account for arbitrary $\kappa$, $u_2$, and $u_3$ \cite{Wang2020}.  In Fig.~\ref{fig:transmission}(a), we show a contour plot of $r_3$ as a function of the real and imaginary parts of the three-body interaction $u_3$.  Interestingly, $r_3$ does not increase arbitrarily as the three-body loss rate is increased, but instead has a maximum value at  $\textrm{Im}(u_3) \sim \Gamma$.  In  Fig.~\ref{fig:transmission}(b), we show several cuts at different values of $\textrm{Im}(u_3)$ near the maximum in $r_3$.    The appearance of a maximum in $r_3$ is attributable to a type of quantum Zeno effect, whereby too large a value of three-body loss blocks the photons from entering the cavity, reducing the overall amount of loss in the system.
	
	{\it Outlook.---}In this Letter,  we showed the existence of a parameter regime for Rydberg polaritons where three-body loss can be resonantly enhanced.  We focused on  dissipative dynamics because, for currently accessible experimental parameters~\cite{Sommer2016,Schine2016,Jia2018,Clark2019b}, the dissipative interactions can be strongly enhanced by working close to the resonance.  Through further experimental improvements and by tuning slightly away from the resonance, one could also operate in a regime where dispersive three-body interactions are strongly enhanced, which is an interesting direction for future studies.
	We would like to stress that although our results are based on a perturbative expansion, this does not mean the interactions are weak.
	On the contrary, the asymptotic expansion in $r_b/L$ means that our results hold for arbitrary optical depths and can give rise to strong effects on the correlations between few photons~\cite{Murray2016}.
	To efficiently study the many-body regime in quasi-1D geometries, one can apply matrix-product-state methods \cite{Manzoni2017,Bienias2020c}. The extension of the presented work to free space is another important direction to explore. Our work clearly demonstrates the possibilities offered by Rydberg-EIT to tune the properties of  multi-body interactions.  This motivates further exploration of possible interactions, which might give rise to different exotic phases of matter \cite{Grass2018,Bienias2020a}.
	
	\acknowledgments{
	{\it Acknowledgments.---}M.K., Y.W., P.B., and A.V.G.\ acknowledge support by ARL CDQI, AFOSR, ARO MURI, DoE ASCR Quantum Testbed Pathfinder program (award No.\ DE-SC0019040), U.S.\ Department of Energy Award No.\ DE-SC0019449, DoE ASCR Accelerated Research in Quantum Computing program (award No.\ DE-SC0020312), NSF PFCQC program, and NSF PFC at JQI. 
	M.K. also acknowledges financial support from the Foundation for Polish Science within the First Team program co-financed by the European Union under the European Regional Development Fund. H.P.B acknowledges funding from the European Research Council
(ERC), Grant agreement No. 681208.}
	
\bibliography{library.bib}

%Control: production of eprint (1) enabled
\begin{thebibliography}{64}%
\makeatletter
\providecommand \@ifxundefined [1]{%
 \@ifx{#1\undefined}
}%
\providecommand \@ifnum [1]{%
 \ifnum #1\expandafter \@firstoftwo
 \else \expandafter \@secondoftwo
 \fi
}%
\providecommand \@ifx [1]{%
 \ifx #1\expandafter \@firstoftwo
 \else \expandafter \@secondoftwo
 \fi
}%
\providecommand \natexlab [1]{#1}%
\providecommand \enquote  [1]{``#1''}%
\providecommand \bibnamefont  [1]{#1}%
\providecommand \bibfnamefont [1]{#1}%
\providecommand \citenamefont [1]{#1}%
\providecommand \href@noop [0]{\@secondoftwo}%
\providecommand \href [0]{\begingroup \@sanitize@url \@href}%
\providecommand \@href[1]{\@@startlink{#1}\@@href}%
\providecommand \@@href[1]{\endgroup#1\@@endlink}%
\providecommand \@sanitize@url [0]{\catcode `\\12\catcode `\$12\catcode
  `\&12\catcode `\#12\catcode `\^12\catcode `\_12\catcode `\%12\relax}%
\providecommand \@@startlink[1]{}%
\providecommand \@@endlink[0]{}%
\providecommand \url  [0]{\begingroup\@sanitize@url \@url }%
\providecommand \@url [1]{\endgroup\@href {#1}{\urlprefix }}%
\providecommand \urlprefix  [0]{URL }%
\providecommand \Eprint [0]{\href }%
\providecommand \doibase [0]{https://doi.org/}%
\providecommand \selectlanguage [0]{\@gobble}%
\providecommand \bibinfo  [0]{\@secondoftwo}%
\providecommand \bibfield  [0]{\@secondoftwo}%
\providecommand \translation [1]{[#1]}%
\providecommand \BibitemOpen [0]{}%
\providecommand \bibitemStop [0]{}%
\providecommand \bibitemNoStop [0]{.\EOS\space}%
\providecommand \EOS [0]{\spacefactor3000\relax}%
\providecommand \BibitemShut  [1]{\csname bibitem#1\endcsname}%
\let\auto@bib@innerbib\@empty
%</preamble>
\bibitem [{\citenamefont {Chang}\ \emph {et~al.}(2014)\citenamefont {Chang},
  \citenamefont {Vuleti{\'{c}}},\ and\ \citenamefont {Lukin}}]{Chang2014}%
  \BibitemOpen
  \bibfield  {author} {\bibinfo {author} {\bibfnamefont {D.~E.}\ \bibnamefont
  {Chang}}, \bibinfo {author} {\bibfnamefont {V.}~\bibnamefont
  {Vuleti{\'{c}}}},\ and\ \bibinfo {author} {\bibfnamefont {M.~D.}\
  \bibnamefont {Lukin}},\ }\bibfield  {title} {\bibinfo {title} {{Quantum
  nonlinear optics - Photon by photon}},\ }\href
  {https://doi.org/10.1038/nphoton.2014.192} {\bibfield  {journal} {\bibinfo
  {journal} {Nat. Photonics}\ }\textbf {\bibinfo {volume} {8}},\ \bibinfo
  {pages} {685} (\bibinfo {year} {2014})}\BibitemShut {NoStop}%
\bibitem [{\citenamefont {Kimble}(2008)}]{Kimble2008}%
  \BibitemOpen
  \bibfield  {author} {\bibinfo {author} {\bibfnamefont {H.~J.}\ \bibnamefont
  {Kimble}},\ }\bibfield  {title} {\bibinfo {title} {{The quantum internet}},\
  }\href {https://doi.org/10.1038/nature07127} {\bibfield  {journal} {\bibinfo
  {journal} {Nature}\ }\textbf {\bibinfo {volume} {453}},\ \bibinfo {pages}
  {1023} (\bibinfo {year} {2008})}\BibitemShut {NoStop}%
\bibitem [{\citenamefont {Carusotto}\ and\ \citenamefont
  {Ciuti}(2013)}]{Carusotto2013}%
  \BibitemOpen
  \bibfield  {author} {\bibinfo {author} {\bibfnamefont {I.}~\bibnamefont
  {Carusotto}}\ and\ \bibinfo {author} {\bibfnamefont {C.}~\bibnamefont
  {Ciuti}},\ }\bibfield  {title} {\bibinfo {title} {Quantum fluids of light},\
  }\href {https://doi.org/10.1103/RevModPhys.85.299} {\bibfield  {journal}
  {\bibinfo  {journal} {Rev. Mod. Phys.}\ }\textbf {\bibinfo {volume} {85}},\
  \bibinfo {pages} {299} (\bibinfo {year} {2013})}\BibitemShut {NoStop}%
\bibitem [{\citenamefont {Maghrebi}\ \emph {et~al.}(2015)\citenamefont
  {Maghrebi}, \citenamefont {Yao}, \citenamefont {Hafezi}, \citenamefont
  {Pohl}, \citenamefont {Firstenberg},\ and\ \citenamefont
  {Gorshkov}}]{Maghrebi2015}%
  \BibitemOpen
  \bibfield  {author} {\bibinfo {author} {\bibfnamefont {M.~F.}\ \bibnamefont
  {Maghrebi}}, \bibinfo {author} {\bibfnamefont {N.~Y.}\ \bibnamefont {Yao}},
  \bibinfo {author} {\bibfnamefont {M.}~\bibnamefont {Hafezi}}, \bibinfo
  {author} {\bibfnamefont {T.}~\bibnamefont {Pohl}}, \bibinfo {author}
  {\bibfnamefont {O.}~\bibnamefont {Firstenberg}},\ and\ \bibinfo {author}
  {\bibfnamefont {A.~V.}\ \bibnamefont {Gorshkov}},\ }\bibfield  {title}
  {\bibinfo {title} {{Fractional quantum Hall states of Rydberg polaritons}},\
  }\href {https://doi.org/10.1103/PhysRevA.91.033838} {\bibfield  {journal}
  {\bibinfo  {journal} {Phys. Rev. A}\ }\textbf {\bibinfo {volume} {91}},\
  \bibinfo {pages} {033838} (\bibinfo {year} {2015})}\BibitemShut {NoStop}%
\bibitem [{\citenamefont {Otterbach}\ \emph {et~al.}(2013)\citenamefont
  {Otterbach}, \citenamefont {Moos}, \citenamefont {Muth},\ and\ \citenamefont
  {Fleischhauer}}]{Otterbach2013}%
  \BibitemOpen
  \bibfield  {author} {\bibinfo {author} {\bibfnamefont {J.}~\bibnamefont
  {Otterbach}}, \bibinfo {author} {\bibfnamefont {M.}~\bibnamefont {Moos}},
  \bibinfo {author} {\bibfnamefont {D.}~\bibnamefont {Muth}},\ and\ \bibinfo
  {author} {\bibfnamefont {M.}~\bibnamefont {Fleischhauer}},\ }\bibfield
  {title} {\bibinfo {title} {{Wigner crystallization of single photons in cold
  Rydberg ensembles}},\ }\href {https://doi.org/10.1103/PhysRevLett.111.113001}
  {\bibfield  {journal} {\bibinfo  {journal} {Phys. Rev. Lett.}\ }\textbf
  {\bibinfo {volume} {111}},\ \bibinfo {pages} {113001} (\bibinfo {year}
  {2013})}\BibitemShut {NoStop}%
\bibitem [{\citenamefont {Efimov}(1970)}]{Efimov1970}%
  \BibitemOpen
  \bibfield  {author} {\bibinfo {author} {\bibfnamefont {V.}~\bibnamefont
  {Efimov}},\ }\bibfield  {title} {\bibinfo {title} {Energy levels arising from
  resonant two-body forces in a three-body system},\ }\href
  {https://doi.org/10.1016/0370-2693(70)90349-7} {\bibfield  {journal}
  {\bibinfo  {journal} {Phys. Lett. B}\ }\textbf {\bibinfo {volume} {33}},\
  \bibinfo {pages} {563 } (\bibinfo {year} {1970})}\BibitemShut {NoStop}%
\bibitem [{\citenamefont {Brown}\ and\ \citenamefont
  {Green}(1969)}]{Brown1969}%
  \BibitemOpen
  \bibfield  {author} {\bibinfo {author} {\bibfnamefont {G.}~\bibnamefont
  {Brown}}\ and\ \bibinfo {author} {\bibfnamefont {A.}~\bibnamefont {Green}},\
  }\bibfield  {title} {\bibinfo {title} {Three-body forces in nuclear matter},\
  }\href {https://doi.org/10.1016/0375-9474(69)90068-2} {\bibfield  {journal}
  {\bibinfo  {journal} {Nucl. Phys. A}\ }\textbf {\bibinfo {volume} {137}},\
  \bibinfo {pages} {1 } (\bibinfo {year} {1969})}\BibitemShut {NoStop}%
\bibitem [{\citenamefont {Steiner}\ and\ \citenamefont
  {Gandolfi}(2012)}]{Steiner2012}%
  \BibitemOpen
  \bibfield  {author} {\bibinfo {author} {\bibfnamefont {A.~W.}\ \bibnamefont
  {Steiner}}\ and\ \bibinfo {author} {\bibfnamefont {S.}~\bibnamefont
  {Gandolfi}},\ }\bibfield  {title} {\bibinfo {title} {{Connecting neutron star
  observations to three-body forces in neutron matter and to the nuclear
  symmetry energy}},\ }\href {https://doi.org/10.1103/PhysRevLett.108.081102}
  {\bibfield  {journal} {\bibinfo  {journal} {Phys. Rev. Lett.}\ }\textbf
  {\bibinfo {volume} {108}},\ \bibinfo {pages} {081102} (\bibinfo {year}
  {2012})}\BibitemShut {NoStop}%
\bibitem [{\citenamefont {Moore}\ and\ \citenamefont {Read}(1991)}]{Moore1991}%
  \BibitemOpen
  \bibfield  {author} {\bibinfo {author} {\bibfnamefont {G.}~\bibnamefont
  {Moore}}\ and\ \bibinfo {author} {\bibfnamefont {N.}~\bibnamefont {Read}},\
  }\bibfield  {title} {\bibinfo {title} {Nonabelions in the fractional quantum
  hall effect},\ }\href {https://doi.org/10.1016/0550-3213(91)90407-O}
  {\bibfield  {journal} {\bibinfo  {journal} {Nucl. Phys. B}\ }\textbf
  {\bibinfo {volume} {360}},\ \bibinfo {pages} {362 } (\bibinfo {year}
  {1991})}\BibitemShut {NoStop}%
\bibitem [{\citenamefont {Fleischhauer}(2005)}]{Fleischhauer2005}%
  \BibitemOpen
  \bibfield  {author} {\bibinfo {author} {\bibfnamefont {M.}~\bibnamefont
  {Fleischhauer}},\ }\bibfield  {title} {\bibinfo {title} {{Electromagnetically
  induced transparency: Optics in coherent media}},\ }\href
  {http://rmp.aps.org/abstract/RMP/v77/i2/p633_1} {\bibfield  {journal}
  {\bibinfo  {journal} {Rev. Mod. Phys.}\ }\textbf {\bibinfo {volume} {77}},\
  \bibinfo {pages} {633} (\bibinfo {year} {2005})}\BibitemShut {NoStop}%
\bibitem [{\citenamefont {Dudin}\ and\ \citenamefont
  {Kuzmich}(2012)}]{Dudin2012}%
  \BibitemOpen
  \bibfield  {author} {\bibinfo {author} {\bibfnamefont {Y.~O.}\ \bibnamefont
  {Dudin}}\ and\ \bibinfo {author} {\bibfnamefont {A.}~\bibnamefont
  {Kuzmich}},\ }\bibfield  {title} {\bibinfo {title} {{Strongly interacting
  Rydberg excitations of a cold atomic gas}},\ }\href
  {https://doi.org/10.1126/science.1217901} {\bibfield  {journal} {\bibinfo
  {journal} {Science}\ }\textbf {\bibinfo {volume} {336}},\ \bibinfo {pages}
  {887} (\bibinfo {year} {2012})}\BibitemShut {NoStop}%
\bibitem [{\citenamefont {Peyronel}\ \emph {et~al.}(2012)\citenamefont
  {Peyronel}, \citenamefont {Firstenberg}, \citenamefont {Liang}, \citenamefont
  {Hofferberth}, \citenamefont {Gorshkov}, \citenamefont {Pohl}, \citenamefont
  {Lukin},\ and\ \citenamefont {Vuleti{\'{c}}}}]{Peyronel2012}%
  \BibitemOpen
  \bibfield  {author} {\bibinfo {author} {\bibfnamefont {T.}~\bibnamefont
  {Peyronel}}, \bibinfo {author} {\bibfnamefont {O.}~\bibnamefont
  {Firstenberg}}, \bibinfo {author} {\bibfnamefont {Q.~Y.}\ \bibnamefont
  {Liang}}, \bibinfo {author} {\bibfnamefont {S.}~\bibnamefont {Hofferberth}},
  \bibinfo {author} {\bibfnamefont {A.~V.}\ \bibnamefont {Gorshkov}}, \bibinfo
  {author} {\bibfnamefont {T.}~\bibnamefont {Pohl}}, \bibinfo {author}
  {\bibfnamefont {M.~D.}\ \bibnamefont {Lukin}},\ and\ \bibinfo {author}
  {\bibfnamefont {V.}~\bibnamefont {Vuleti{\'{c}}}},\ }\bibfield  {title}
  {\bibinfo {title} {{Quantum nonlinear optics with single photons enabled by
  strongly interacting atoms}},\ }\href {https://doi.org/10.1038/nature11361}
  {\bibfield  {journal} {\bibinfo  {journal} {Nature}\ }\textbf {\bibinfo
  {volume} {488}},\ \bibinfo {pages} {57} (\bibinfo {year} {2012})}\BibitemShut
  {NoStop}%
\bibitem [{\citenamefont {Maxwell}\ \emph {et~al.}(2013)\citenamefont
  {Maxwell}, \citenamefont {Szwer}, \citenamefont {Paredes-Barato},
  \citenamefont {Busche}, \citenamefont {Pritchard}, \citenamefont {Gauguet},
  \citenamefont {Weatherill}, \citenamefont {Jones},\ and\ \citenamefont
  {Adams}}]{Maxwell2013}%
  \BibitemOpen
  \bibfield  {author} {\bibinfo {author} {\bibfnamefont {D.}~\bibnamefont
  {Maxwell}}, \bibinfo {author} {\bibfnamefont {D.~J.}\ \bibnamefont {Szwer}},
  \bibinfo {author} {\bibfnamefont {D.}~\bibnamefont {Paredes-Barato}},
  \bibinfo {author} {\bibfnamefont {H.}~\bibnamefont {Busche}}, \bibinfo
  {author} {\bibfnamefont {J.~D.}\ \bibnamefont {Pritchard}}, \bibinfo {author}
  {\bibfnamefont {A.}~\bibnamefont {Gauguet}}, \bibinfo {author} {\bibfnamefont
  {K.~J.}\ \bibnamefont {Weatherill}}, \bibinfo {author} {\bibfnamefont
  {M.~P.~A.}\ \bibnamefont {Jones}},\ and\ \bibinfo {author} {\bibfnamefont
  {C.~S.}\ \bibnamefont {Adams}},\ }\bibfield  {title} {\bibinfo {title}
  {{Storage and control of optical photons using Rydberg polaritons}},\ }\href
  {https://doi.org/10.1103/PhysRevLett.110.103001} {\bibfield  {journal}
  {\bibinfo  {journal} {Phys. Rev. Lett.}\ }\textbf {\bibinfo {volume} {110}},\
  \bibinfo {pages} {103001} (\bibinfo {year} {2013})}\BibitemShut {NoStop}%
\bibitem [{\citenamefont {Stanojevic}\ \emph {et~al.}(2013)\citenamefont
  {Stanojevic}, \citenamefont {Parigi}, \citenamefont {Bimbard}, \citenamefont
  {Ourjoumtsev},\ and\ \citenamefont {Grangier}}]{Stanojevic2013}%
  \BibitemOpen
  \bibfield  {author} {\bibinfo {author} {\bibfnamefont {J.}~\bibnamefont
  {Stanojevic}}, \bibinfo {author} {\bibfnamefont {V.}~\bibnamefont {Parigi}},
  \bibinfo {author} {\bibfnamefont {E.}~\bibnamefont {Bimbard}}, \bibinfo
  {author} {\bibfnamefont {A.}~\bibnamefont {Ourjoumtsev}},\ and\ \bibinfo
  {author} {\bibfnamefont {P.}~\bibnamefont {Grangier}},\ }\bibfield  {title}
  {\bibinfo {title} {{Dispersive optical nonlinearities in a Rydberg
  electromagnetically-induced-transparency medium}},\ }\href
  {https://doi.org/10.1103/PhysRevA.88.053845} {\bibfield  {journal} {\bibinfo
  {journal} {Phys. Rev. A}\ }\textbf {\bibinfo {volume} {88}},\ \bibinfo
  {pages} {053845} (\bibinfo {year} {2013})}\BibitemShut {NoStop}%
\bibitem [{\citenamefont {Li}\ \emph {et~al.}(2016)\citenamefont {Li},
  \citenamefont {Snyder}, \citenamefont {Pelaschier}, \citenamefont {Huang},
  \citenamefont {Niranjan}, \citenamefont {Duncan}, \citenamefont {Rupp},
  \citenamefont {M{\"{u}}ller},\ and\ \citenamefont {Burke}}]{Li2016}%
  \BibitemOpen
  \bibfield  {author} {\bibinfo {author} {\bibfnamefont {L.}~\bibnamefont
  {Li}}, \bibinfo {author} {\bibfnamefont {J.~C.}\ \bibnamefont {Snyder}},
  \bibinfo {author} {\bibfnamefont {I.~M.}\ \bibnamefont {Pelaschier}},
  \bibinfo {author} {\bibfnamefont {J.}~\bibnamefont {Huang}}, \bibinfo
  {author} {\bibfnamefont {U.~N.}\ \bibnamefont {Niranjan}}, \bibinfo {author}
  {\bibfnamefont {P.}~\bibnamefont {Duncan}}, \bibinfo {author} {\bibfnamefont
  {M.}~\bibnamefont {Rupp}}, \bibinfo {author} {\bibfnamefont {K.~R.}\
  \bibnamefont {M{\"{u}}ller}},\ and\ \bibinfo {author} {\bibfnamefont
  {K.}~\bibnamefont {Burke}},\ }\bibfield  {title} {\bibinfo {title}
  {{Understanding machine-learned density functionals}},\ }\href
  {https://doi.org/10.1002/qua.25040} {\bibfield  {journal} {\bibinfo
  {journal} {Int. J. Quantum Chem.}\ }\textbf {\bibinfo {volume} {116}},\
  \bibinfo {pages} {819} (\bibinfo {year} {2016})}\BibitemShut {NoStop}%
\bibitem [{\citenamefont {Gorniaczyk}\ \emph {et~al.}(2014)\citenamefont
  {Gorniaczyk}, \citenamefont {Tresp}, \citenamefont {Schmidt}, \citenamefont
  {Fedder},\ and\ \citenamefont {Hofferberth}}]{Gorniaczyk2014}%
  \BibitemOpen
  \bibfield  {author} {\bibinfo {author} {\bibfnamefont {H.}~\bibnamefont
  {Gorniaczyk}}, \bibinfo {author} {\bibfnamefont {C.}~\bibnamefont {Tresp}},
  \bibinfo {author} {\bibfnamefont {J.}~\bibnamefont {Schmidt}}, \bibinfo
  {author} {\bibfnamefont {H.}~\bibnamefont {Fedder}},\ and\ \bibinfo {author}
  {\bibfnamefont {S.}~\bibnamefont {Hofferberth}},\ }\bibfield  {title}
  {\bibinfo {title} {{Single-photon transistor mediated by interstate Rydberg
  interactions}},\ }\href {https://doi.org/10.1103/PhysRevLett.113.053601}
  {\bibfield  {journal} {\bibinfo  {journal} {Phys. Rev. Lett.}\ }\textbf
  {\bibinfo {volume} {113}},\ \bibinfo {pages} {053601} (\bibinfo {year}
  {2014})}\BibitemShut {NoStop}%
\bibitem [{\citenamefont {Tiarks}\ \emph {et~al.}(2014)\citenamefont {Tiarks},
  \citenamefont {Baur}, \citenamefont {Schneider}, \citenamefont {D{\"{u}}rr},\
  and\ \citenamefont {Rempe}}]{Tiarks2014}%
  \BibitemOpen
  \bibfield  {author} {\bibinfo {author} {\bibfnamefont {D.}~\bibnamefont
  {Tiarks}}, \bibinfo {author} {\bibfnamefont {S.}~\bibnamefont {Baur}},
  \bibinfo {author} {\bibfnamefont {K.}~\bibnamefont {Schneider}}, \bibinfo
  {author} {\bibfnamefont {S.}~\bibnamefont {D{\"{u}}rr}},\ and\ \bibinfo
  {author} {\bibfnamefont {G.}~\bibnamefont {Rempe}},\ }\bibfield  {title}
  {\bibinfo {title} {{Single-photon transistor using a F{\"{o}}rster
  resonance}},\ }\href {https://doi.org/10.1103/PhysRevLett.113.053602}
  {\bibfield  {journal} {\bibinfo  {journal} {Phys. Rev. Lett.}\ }\textbf
  {\bibinfo {volume} {113}},\ \bibinfo {pages} {053602} (\bibinfo {year}
  {2014})}\BibitemShut {NoStop}%
\bibitem [{\citenamefont {Gorniaczyk}\ \emph {et~al.}(2016)\citenamefont
  {Gorniaczyk}, \citenamefont {Tresp}, \citenamefont {Bienias}, \citenamefont
  {Paris-Mandoki}, \citenamefont {Li}, \citenamefont {Mirgorodskiy},
  \citenamefont {B{\"{u}}chler}, \citenamefont {Lesanovsky},\ and\
  \citenamefont {Hofferberth}}]{Gorniaczyk2016}%
  \BibitemOpen
  \bibfield  {author} {\bibinfo {author} {\bibfnamefont {H.}~\bibnamefont
  {Gorniaczyk}}, \bibinfo {author} {\bibfnamefont {C.}~\bibnamefont {Tresp}},
  \bibinfo {author} {\bibfnamefont {P.}~\bibnamefont {Bienias}}, \bibinfo
  {author} {\bibfnamefont {A.}~\bibnamefont {Paris-Mandoki}}, \bibinfo {author}
  {\bibfnamefont {W.}~\bibnamefont {Li}}, \bibinfo {author} {\bibfnamefont
  {I.}~\bibnamefont {Mirgorodskiy}}, \bibinfo {author} {\bibfnamefont {H.~P.}\
  \bibnamefont {B{\"{u}}chler}}, \bibinfo {author} {\bibfnamefont
  {I.}~\bibnamefont {Lesanovsky}},\ and\ \bibinfo {author} {\bibfnamefont
  {S.}~\bibnamefont {Hofferberth}},\ }\bibfield  {title} {\bibinfo {title}
  {{Enhancement of Rydberg-mediated single-photon nonlinearities by
  electrically tuned F{\"{o}}rster resonances}},\ }\href
  {https://doi.org/10.1038/ncomms12480} {\bibfield  {journal} {\bibinfo
  {journal} {Nat. Commun.}\ }\textbf {\bibinfo {volume} {7}},\ \bibinfo {pages}
  {12480} (\bibinfo {year} {2016})}\BibitemShut {NoStop}%
\bibitem [{\citenamefont {Tiarks}\ \emph {et~al.}(2016)\citenamefont {Tiarks},
  \citenamefont {Schmidt}, \citenamefont {Rempe},\ and\ \citenamefont
  {D{\"{u}}rr}}]{Tiarks2016}%
  \BibitemOpen
  \bibfield  {author} {\bibinfo {author} {\bibfnamefont {D.}~\bibnamefont
  {Tiarks}}, \bibinfo {author} {\bibfnamefont {S.}~\bibnamefont {Schmidt}},
  \bibinfo {author} {\bibfnamefont {G.}~\bibnamefont {Rempe}},\ and\ \bibinfo
  {author} {\bibfnamefont {S.}~\bibnamefont {D{\"{u}}rr}},\ }\bibfield  {title}
  {\bibinfo {title} {{Optical $\pi$ phase shift created with a single-photon
  pulse}},\ }\href {https://doi.org/10.1126/sciadv.1600036} {\bibfield
  {journal} {\bibinfo  {journal} {Sci. Adv.}\ }\textbf {\bibinfo {volume}
  {2}},\ \bibinfo {pages} {e1600036} (\bibinfo {year} {2016})}\BibitemShut
  {NoStop}%
\bibitem [{\citenamefont {Thompson}\ \emph {et~al.}(2017)\citenamefont
  {Thompson}, \citenamefont {Nicholson}, \citenamefont {Liang}, \citenamefont
  {Cantu}, \citenamefont {Venkatramani}, \citenamefont {Choi}, \citenamefont
  {Fedorov}, \citenamefont {Viscor}, \citenamefont {Pohl}, \citenamefont
  {Lukin},\ and\ \citenamefont {Vuletic}}]{Thompson2017}%
  \BibitemOpen
  \bibfield  {author} {\bibinfo {author} {\bibfnamefont {J.~D.}\ \bibnamefont
  {Thompson}}, \bibinfo {author} {\bibfnamefont {T.~L.}\ \bibnamefont
  {Nicholson}}, \bibinfo {author} {\bibfnamefont {Q.~Y.}\ \bibnamefont
  {Liang}}, \bibinfo {author} {\bibfnamefont {S.~H.}\ \bibnamefont {Cantu}},
  \bibinfo {author} {\bibfnamefont {A.~V.}\ \bibnamefont {Venkatramani}},
  \bibinfo {author} {\bibfnamefont {S.}~\bibnamefont {Choi}}, \bibinfo {author}
  {\bibfnamefont {I.~A.}\ \bibnamefont {Fedorov}}, \bibinfo {author}
  {\bibfnamefont {D.}~\bibnamefont {Viscor}}, \bibinfo {author} {\bibfnamefont
  {T.}~\bibnamefont {Pohl}}, \bibinfo {author} {\bibfnamefont {M.~D.}\
  \bibnamefont {Lukin}},\ and\ \bibinfo {author} {\bibfnamefont
  {V.}~\bibnamefont {Vuletic}},\ }\bibfield  {title} {\bibinfo {title}
  {{Symmetry-protected collisions between strongly interacting photons}},\
  }\href {https://doi.org/10.1038/nature20823} {\bibfield  {journal} {\bibinfo
  {journal} {Nature}\ }\textbf {\bibinfo {volume} {542}},\ \bibinfo {pages}
  {206} (\bibinfo {year} {2017})}\BibitemShut {NoStop}%
\bibitem [{\citenamefont {Tiarks}\ \emph {et~al.}(2019)\citenamefont {Tiarks},
  \citenamefont {Schmidt-Eberle}, \citenamefont {Stolz}, \citenamefont
  {Rempe},\ and\ \citenamefont {D{\"{u}}rr}}]{Tiarks2019}%
  \BibitemOpen
  \bibfield  {author} {\bibinfo {author} {\bibfnamefont {D.}~\bibnamefont
  {Tiarks}}, \bibinfo {author} {\bibfnamefont {S.}~\bibnamefont
  {Schmidt-Eberle}}, \bibinfo {author} {\bibfnamefont {T.}~\bibnamefont
  {Stolz}}, \bibinfo {author} {\bibfnamefont {G.}~\bibnamefont {Rempe}},\ and\
  \bibinfo {author} {\bibfnamefont {S.}~\bibnamefont {D{\"{u}}rr}},\ }\bibfield
   {title} {\bibinfo {title} {{A photon–photon quantum gate based on Rydberg
  interactions}},\ }\href {https://doi.org/10.1038/s41567-018-0313-7}
  {\bibfield  {journal} {\bibinfo  {journal} {Nat. Phys.}\ }\textbf {\bibinfo
  {volume} {15}},\ \bibinfo {pages} {124} (\bibinfo {year} {2019})}\BibitemShut
  {NoStop}%
\bibitem [{\citenamefont {Firstenberg}\ \emph {et~al.}(2013)\citenamefont
  {Firstenberg}, \citenamefont {Peyronel}, \citenamefont {Liang}, \citenamefont
  {Gorshkov}, \citenamefont {Lukin},\ and\ \citenamefont
  {Vuleti{\'{c}}}}]{Firstenberg2013}%
  \BibitemOpen
  \bibfield  {author} {\bibinfo {author} {\bibfnamefont {O.}~\bibnamefont
  {Firstenberg}}, \bibinfo {author} {\bibfnamefont {T.}~\bibnamefont
  {Peyronel}}, \bibinfo {author} {\bibfnamefont {Q.~Y.}\ \bibnamefont {Liang}},
  \bibinfo {author} {\bibfnamefont {A.~V.}\ \bibnamefont {Gorshkov}}, \bibinfo
  {author} {\bibfnamefont {M.~D.}\ \bibnamefont {Lukin}},\ and\ \bibinfo
  {author} {\bibfnamefont {V.}~\bibnamefont {Vuleti{\'{c}}}},\ }\bibfield
  {title} {\bibinfo {title} {{Attractive photons in a quantum nonlinear
  medium}},\ }\href {https://doi.org/10.1038/nature12512} {\bibfield  {journal}
  {\bibinfo  {journal} {Nature}\ }\textbf {\bibinfo {volume} {502}},\ \bibinfo
  {pages} {71} (\bibinfo {year} {2013})}\BibitemShut {NoStop}%
\bibitem [{\citenamefont {Liang}\ \emph {et~al.}(2018)\citenamefont {Liang},
  \citenamefont {Venkatramani}, \citenamefont {Cantu}, \citenamefont
  {Nicholson}, \citenamefont {Gullans}, \citenamefont {Gorshkov}, \citenamefont
  {Thompson}, \citenamefont {Chin}, \citenamefont {Lukin},\ and\ \citenamefont
  {Vuleti{\'{c}}}}]{Liang2018}%
  \BibitemOpen
  \bibfield  {author} {\bibinfo {author} {\bibfnamefont {Q.-Y.}\ \bibnamefont
  {Liang}}, \bibinfo {author} {\bibfnamefont {A.~V.}\ \bibnamefont
  {Venkatramani}}, \bibinfo {author} {\bibfnamefont {S.~H.}\ \bibnamefont
  {Cantu}}, \bibinfo {author} {\bibfnamefont {T.~L.}\ \bibnamefont
  {Nicholson}}, \bibinfo {author} {\bibfnamefont {M.~J.}\ \bibnamefont
  {Gullans}}, \bibinfo {author} {\bibfnamefont {A.~V.}\ \bibnamefont
  {Gorshkov}}, \bibinfo {author} {\bibfnamefont {J.~D.}\ \bibnamefont
  {Thompson}}, \bibinfo {author} {\bibfnamefont {C.}~\bibnamefont {Chin}},
  \bibinfo {author} {\bibfnamefont {M.~D.}\ \bibnamefont {Lukin}},\ and\
  \bibinfo {author} {\bibfnamefont {V.}~\bibnamefont {Vuleti{\'{c}}}},\
  }\bibfield  {title} {\bibinfo {title} {{Observation of three-photon bound
  states in a quantum nonlinear medium}},\ }\href
  {https://doi.org/10.1126/science.aao7293} {\bibfield  {journal} {\bibinfo
  {journal} {Science}\ }\textbf {\bibinfo {volume} {359}},\ \bibinfo {pages}
  {783} (\bibinfo {year} {2018})}\BibitemShut {NoStop}%
\bibitem [{\citenamefont {Stiesdal}\ \emph {et~al.}(2018)\citenamefont
  {Stiesdal}, \citenamefont {Kumlin}, \citenamefont {Kleinbeck}, \citenamefont
  {Lunt}, \citenamefont {Braun}, \citenamefont {Paris-Mandoki}, \citenamefont
  {Tresp}, \citenamefont {B{\"{u}}chler},\ and\ \citenamefont
  {Hofferberth}}]{Stiesdal2018}%
  \BibitemOpen
  \bibfield  {author} {\bibinfo {author} {\bibfnamefont {N.}~\bibnamefont
  {Stiesdal}}, \bibinfo {author} {\bibfnamefont {J.}~\bibnamefont {Kumlin}},
  \bibinfo {author} {\bibfnamefont {K.}~\bibnamefont {Kleinbeck}}, \bibinfo
  {author} {\bibfnamefont {P.}~\bibnamefont {Lunt}}, \bibinfo {author}
  {\bibfnamefont {C.}~\bibnamefont {Braun}}, \bibinfo {author} {\bibfnamefont
  {A.}~\bibnamefont {Paris-Mandoki}}, \bibinfo {author} {\bibfnamefont
  {C.}~\bibnamefont {Tresp}}, \bibinfo {author} {\bibfnamefont {H.~P.}\
  \bibnamefont {B{\"{u}}chler}},\ and\ \bibinfo {author} {\bibfnamefont
  {S.}~\bibnamefont {Hofferberth}},\ }\bibfield  {title} {\bibinfo {title}
  {{Observation of Three-Body Correlations for Photons Coupled to a Rydberg
  Superatom}},\ }\href {https://doi.org/10.1103/PhysRevLett.121.103601}
  {\bibfield  {journal} {\bibinfo  {journal} {Phys. Rev. Lett.}\ }\textbf
  {\bibinfo {volume} {121}},\ \bibinfo {pages} {103601} (\bibinfo {year}
  {2018})}\BibitemShut {NoStop}%
\bibitem [{\citenamefont {Sommer}\ and\ \citenamefont
  {Simon}(2016)}]{Sommer2016}%
  \BibitemOpen
  \bibfield  {author} {\bibinfo {author} {\bibfnamefont {A.}~\bibnamefont
  {Sommer}}\ and\ \bibinfo {author} {\bibfnamefont {J.}~\bibnamefont {Simon}},\
  }\bibfield  {title} {\bibinfo {title} {{Engineering photonic Floquet
  Hamiltonians through Fabry-P{\'{e}}rot resonators}},\ }\href
  {https://doi.org/10.1088/1367-2630/18/3/035008} {\bibfield  {journal}
  {\bibinfo  {journal} {New J. Phys.}\ }\textbf {\bibinfo {volume} {18}},\
  \bibinfo {pages} {35008} (\bibinfo {year} {2016})}\BibitemShut {NoStop}%
\bibitem [{\citenamefont {Schine}\ \emph {et~al.}(2016)\citenamefont {Schine},
  \citenamefont {Ryou}, \citenamefont {Gromov}, \citenamefont {Sommer},\ and\
  \citenamefont {Simon}}]{Schine2016}%
  \BibitemOpen
  \bibfield  {author} {\bibinfo {author} {\bibfnamefont {N.}~\bibnamefont
  {Schine}}, \bibinfo {author} {\bibfnamefont {A.}~\bibnamefont {Ryou}},
  \bibinfo {author} {\bibfnamefont {A.}~\bibnamefont {Gromov}}, \bibinfo
  {author} {\bibfnamefont {A.}~\bibnamefont {Sommer}},\ and\ \bibinfo {author}
  {\bibfnamefont {J.}~\bibnamefont {Simon}},\ }\bibfield  {title} {\bibinfo
  {title} {{Synthetic Landau levels for photons}},\ }\href
  {https://doi.org/10.1038/nature} {\bibfield  {journal} {\bibinfo  {journal}
  {Nature}\ }\textbf {\bibinfo {volume} {534}},\ \bibinfo {pages} {671}
  (\bibinfo {year} {2016})}\BibitemShut {NoStop}%
\bibitem [{\citenamefont {Jia}\ \emph {et~al.}(2018)\citenamefont {Jia},
  \citenamefont {Schine}, \citenamefont {Georgakopoulos}, \citenamefont {Ryou},
  \citenamefont {Clark}, \citenamefont {Sommer},\ and\ \citenamefont
  {Simon}}]{Jia2018}%
  \BibitemOpen
  \bibfield  {author} {\bibinfo {author} {\bibfnamefont {N.}~\bibnamefont
  {Jia}}, \bibinfo {author} {\bibfnamefont {N.}~\bibnamefont {Schine}},
  \bibinfo {author} {\bibfnamefont {A.}~\bibnamefont {Georgakopoulos}},
  \bibinfo {author} {\bibfnamefont {A.}~\bibnamefont {Ryou}}, \bibinfo {author}
  {\bibfnamefont {L.~W.}\ \bibnamefont {Clark}}, \bibinfo {author}
  {\bibfnamefont {A.}~\bibnamefont {Sommer}},\ and\ \bibinfo {author}
  {\bibfnamefont {J.}~\bibnamefont {Simon}},\ }\bibfield  {title} {\bibinfo
  {title} {{A strongly interacting polaritonic quantum dot}},\ }\href
  {https://doi.org/10.1038/s41567-018-0071-6} {\bibfield  {journal} {\bibinfo
  {journal} {Nat. Phys.}\ }\textbf {\bibinfo {volume} {14}},\ \bibinfo {pages}
  {550} (\bibinfo {year} {2018})}\BibitemShut {NoStop}%
\bibitem [{\citenamefont {Clark}\ \emph {et~al.}(2019)\citenamefont {Clark},
  \citenamefont {Jia}, \citenamefont {Schine}, \citenamefont {Baum},
  \citenamefont {Georgakopoulos},\ and\ \citenamefont {Simon}}]{Clark2019b}%
  \BibitemOpen
  \bibfield  {author} {\bibinfo {author} {\bibfnamefont {L.~W.}\ \bibnamefont
  {Clark}}, \bibinfo {author} {\bibfnamefont {N.}~\bibnamefont {Jia}}, \bibinfo
  {author} {\bibfnamefont {N.}~\bibnamefont {Schine}}, \bibinfo {author}
  {\bibfnamefont {C.}~\bibnamefont {Baum}}, \bibinfo {author} {\bibfnamefont
  {A.}~\bibnamefont {Georgakopoulos}},\ and\ \bibinfo {author} {\bibfnamefont
  {J.}~\bibnamefont {Simon}},\ }\bibfield  {title} {\bibinfo {title}
  {{Interacting Floquet polaritons}},\ }\href
  {https://doi.org/10.1038/s41586-019-1354-5} {\bibfield  {journal} {\bibinfo
  {journal} {Nature}\ }\textbf {\bibinfo {volume} {532}},\ \bibinfo {pages}
  {571} (\bibinfo {year} {2019})}\BibitemShut {NoStop}%
\bibitem [{\citenamefont {Cantu}\ \emph {et~al.}(2020)\citenamefont {Cantu},
  \citenamefont {Venkatramani}, \citenamefont {Xu}, \citenamefont {Zhou},
  \citenamefont {Jelenkovi{\'{c}}}, \citenamefont {Lukin},\ and\ \citenamefont
  {Vuleti{\'{c}}}}]{Cantu2020}%
  \BibitemOpen
  \bibfield  {author} {\bibinfo {author} {\bibfnamefont {S.~H.}\ \bibnamefont
  {Cantu}}, \bibinfo {author} {\bibfnamefont {A.~V.}\ \bibnamefont
  {Venkatramani}}, \bibinfo {author} {\bibfnamefont {W.}~\bibnamefont {Xu}},
  \bibinfo {author} {\bibfnamefont {L.}~\bibnamefont {Zhou}}, \bibinfo {author}
  {\bibfnamefont {B.}~\bibnamefont {Jelenkovi{\'{c}}}}, \bibinfo {author}
  {\bibfnamefont {M.~D.}\ \bibnamefont {Lukin}},\ and\ \bibinfo {author}
  {\bibfnamefont {V.}~\bibnamefont {Vuleti{\'{c}}}},\ }\bibfield  {title}
  {\bibinfo {title} {{Repulsive photons in a quantum nonlinear medium}},\
  }\href {https://doi.org/10.1038/s41567-020-0917-6} {\bibfield  {journal}
  {\bibinfo  {journal} {Nat. Phys.}\ }\textbf {\bibinfo {volume} {16}},\
  \bibinfo {pages} {921} (\bibinfo {year} {2020})}\BibitemShut {NoStop}%
\bibitem [{\citenamefont {Jachymski}\ \emph {et~al.}(2016)\citenamefont
  {Jachymski}, \citenamefont {Bienias},\ and\ \citenamefont
  {B{\"{u}}chler}}]{Jachymski2016}%
  \BibitemOpen
  \bibfield  {author} {\bibinfo {author} {\bibfnamefont {K.}~\bibnamefont
  {Jachymski}}, \bibinfo {author} {\bibfnamefont {P.}~\bibnamefont {Bienias}},\
  and\ \bibinfo {author} {\bibfnamefont {H.~P.}\ \bibnamefont
  {B{\"{u}}chler}},\ }\bibfield  {title} {\bibinfo {title} {{Three-body
  interactions of slow light Rydberg polaritons}},\ }\href
  {https://doi.org/10.1103/PhysRevLett.117.053601} {\bibfield  {journal}
  {\bibinfo  {journal} {Phys. Rev. Lett.}\ }\textbf {\bibinfo {volume} {117}},\
  \bibinfo {pages} {053601} (\bibinfo {year} {2016})}\BibitemShut {NoStop}%
\bibitem [{\citenamefont {Gullans}\ \emph {et~al.}(2016)\citenamefont
  {Gullans}, \citenamefont {Thompson}, \citenamefont {Wang}, \citenamefont
  {Liang}, \citenamefont {Vuleti{\'{c}}}, \citenamefont {Lukin},\ and\
  \citenamefont {Gorshkov}}]{Gullans2016}%
  \BibitemOpen
  \bibfield  {author} {\bibinfo {author} {\bibfnamefont {M.~J.}\ \bibnamefont
  {Gullans}}, \bibinfo {author} {\bibfnamefont {J.~D.}\ \bibnamefont
  {Thompson}}, \bibinfo {author} {\bibfnamefont {Y.}~\bibnamefont {Wang}},
  \bibinfo {author} {\bibfnamefont {Q.~Y.}\ \bibnamefont {Liang}}, \bibinfo
  {author} {\bibfnamefont {V.}~\bibnamefont {Vuleti{\'{c}}}}, \bibinfo {author}
  {\bibfnamefont {M.~D.}\ \bibnamefont {Lukin}},\ and\ \bibinfo {author}
  {\bibfnamefont {A.~V.}\ \bibnamefont {Gorshkov}},\ }\bibfield  {title}
  {\bibinfo {title} {{Effective Field Theory for Rydberg Polaritons}},\ }\href
  {https://doi.org/10.1103/PhysRevLett.117.113601} {\bibfield  {journal}
  {\bibinfo  {journal} {Phys. Rev. Lett.}\ }\textbf {\bibinfo {volume} {117}},\
  \bibinfo {pages} {113601} (\bibinfo {year} {2016})}\BibitemShut {NoStop}%
\bibitem [{\citenamefont {Gullans}\ \emph {et~al.}(2017)\citenamefont
  {Gullans}, \citenamefont {Diehl}, \citenamefont {Rittenhouse}, \citenamefont
  {Ruzic}, \citenamefont {D'Incao}, \citenamefont {Julienne}, \citenamefont
  {Gorshkov},\ and\ \citenamefont {Taylor}}]{Gullans2017}%
  \BibitemOpen
  \bibfield  {author} {\bibinfo {author} {\bibfnamefont {M.~J.}\ \bibnamefont
  {Gullans}}, \bibinfo {author} {\bibfnamefont {S.}~\bibnamefont {Diehl}},
  \bibinfo {author} {\bibfnamefont {S.~T.}\ \bibnamefont {Rittenhouse}},
  \bibinfo {author} {\bibfnamefont {B.~P.}\ \bibnamefont {Ruzic}}, \bibinfo
  {author} {\bibfnamefont {J.~P.}\ \bibnamefont {D'Incao}}, \bibinfo {author}
  {\bibfnamefont {P.}~\bibnamefont {Julienne}}, \bibinfo {author}
  {\bibfnamefont {A.~V.}\ \bibnamefont {Gorshkov}},\ and\ \bibinfo {author}
  {\bibfnamefont {J.~M.}\ \bibnamefont {Taylor}},\ }\bibfield  {title}
  {\bibinfo {title} {{Efimov States of Strongly Interacting Photons}},\ }\href
  {https://doi.org/10.1103/PhysRevLett.119.233601} {\bibfield  {journal}
  {\bibinfo  {journal} {Phys. Rev. Lett.}\ }\textbf {\bibinfo {volume} {119}},\
  \bibinfo {pages} {233601} (\bibinfo {year} {2017})}\BibitemShut {NoStop}%
\bibitem [{\citenamefont {Bienias}\ \emph
  {et~al.}(2020{\natexlab{a}})\citenamefont {Bienias}, \citenamefont {Gullans},
  \citenamefont {Kalinowski}, \citenamefont {Craddock}, \citenamefont
  {Ornelas-Huerta}, \citenamefont {Rolston}, \citenamefont {Porto},\ and\
  \citenamefont {Gorshkov}}]{Bienias2020a}%
  \BibitemOpen
  \bibfield  {author} {\bibinfo {author} {\bibfnamefont {P.}~\bibnamefont
  {Bienias}}, \bibinfo {author} {\bibfnamefont {M.~J.}\ \bibnamefont
  {Gullans}}, \bibinfo {author} {\bibfnamefont {M.}~\bibnamefont {Kalinowski}},
  \bibinfo {author} {\bibfnamefont {A.~N.}\ \bibnamefont {Craddock}}, \bibinfo
  {author} {\bibfnamefont {D.~P.}\ \bibnamefont {Ornelas-Huerta}}, \bibinfo
  {author} {\bibfnamefont {S.~L.}\ \bibnamefont {Rolston}}, \bibinfo {author}
  {\bibfnamefont {J.~V.}\ \bibnamefont {Porto}},\ and\ \bibinfo {author}
  {\bibfnamefont {A.~V.}\ \bibnamefont {Gorshkov}},\ }\bibfield  {title}
  {\bibinfo {title} {Exotic photonic molecules via lennard-jones-like
  potentials},\ }\href {https://doi.org/10.1103/PhysRevLett.125.093601}
  {\bibfield  {journal} {\bibinfo  {journal} {Phys. Rev. Lett.}\ }\textbf
  {\bibinfo {volume} {125}},\ \bibinfo {pages} {093601} (\bibinfo {year}
  {2020}{\natexlab{a}})}\BibitemShut {NoStop}%
\bibitem [{\citenamefont {Gambetta}\ \emph {et~al.}(2020)\citenamefont
  {Gambetta}, \citenamefont {Li}, \citenamefont {Schmidt-Kaler},\ and\
  \citenamefont {Lesanovsky}}]{Gambetta2020}%
  \BibitemOpen
  \bibfield  {author} {\bibinfo {author} {\bibfnamefont {F.~M.}\ \bibnamefont
  {Gambetta}}, \bibinfo {author} {\bibfnamefont {W.}~\bibnamefont {Li}},
  \bibinfo {author} {\bibfnamefont {F.}~\bibnamefont {Schmidt-Kaler}},\ and\
  \bibinfo {author} {\bibfnamefont {I.}~\bibnamefont {Lesanovsky}},\ }\bibfield
   {title} {\bibinfo {title} {{Engineering NonBinary Rydberg Interactions via
  Phonons in an Optical Lattice}},\ }\href
  {https://doi.org/10.1103/physrevlett.124.043402} {\bibfield  {journal}
  {\bibinfo  {journal} {Phys. Rev. Lett.}\ }\textbf {\bibinfo {volume} {124}},\
  \bibinfo {pages} {043402} (\bibinfo {year} {2020})}\BibitemShut {NoStop}%
\bibitem [{\citenamefont {B{\"{u}}chler}\ \emph {et~al.}(2007)\citenamefont
  {B{\"{u}}chler}, \citenamefont {Micheli},\ and\ \citenamefont
  {Zoller}}]{Buchler2007a}%
  \BibitemOpen
  \bibfield  {author} {\bibinfo {author} {\bibfnamefont {H.~P.}\ \bibnamefont
  {B{\"{u}}chler}}, \bibinfo {author} {\bibfnamefont {A.}~\bibnamefont
  {Micheli}},\ and\ \bibinfo {author} {\bibfnamefont {P.}~\bibnamefont
  {Zoller}},\ }\bibfield  {title} {\bibinfo {title} {{Three-body interactions
  with cold polar molecules}},\ }\href {https://doi.org/10.1038/nphys678}
  {\bibfield  {journal} {\bibinfo  {journal} {Nat. Phys.}\ }\textbf {\bibinfo
  {volume} {3}},\ \bibinfo {pages} {726} (\bibinfo {year} {2007})}\BibitemShut
  {NoStop}%
\bibitem [{\citenamefont {Daley}\ \emph {et~al.}(2009)\citenamefont {Daley},
  \citenamefont {Taylor}, \citenamefont {Diehl}, \citenamefont {Baranov},\ and\
  \citenamefont {Zoller}}]{Daley2009}%
  \BibitemOpen
  \bibfield  {author} {\bibinfo {author} {\bibfnamefont {A.~J.}\ \bibnamefont
  {Daley}}, \bibinfo {author} {\bibfnamefont {J.~M.}\ \bibnamefont {Taylor}},
  \bibinfo {author} {\bibfnamefont {S.}~\bibnamefont {Diehl}}, \bibinfo
  {author} {\bibfnamefont {M.}~\bibnamefont {Baranov}},\ and\ \bibinfo {author}
  {\bibfnamefont {P.}~\bibnamefont {Zoller}},\ }\bibfield  {title} {\bibinfo
  {title} {{Atomic three-body loss as a dynamical three-body interaction}},\
  }\href {https://doi.org/10.1103/PhysRevLett.102.040402} {\bibfield  {journal}
  {\bibinfo  {journal} {Phys. Rev. Lett.}\ }\textbf {\bibinfo {volume} {102}},\
  \bibinfo {pages} {040402} (\bibinfo {year} {2009})}\BibitemShut {NoStop}%
\bibitem [{\citenamefont {Johnson}\ \emph {et~al.}(2009)\citenamefont
  {Johnson}, \citenamefont {Tiesinga}, \citenamefont {Porto},\ and\
  \citenamefont {Williams}}]{Johnson2009}%
  \BibitemOpen
  \bibfield  {author} {\bibinfo {author} {\bibfnamefont {P.~R.}\ \bibnamefont
  {Johnson}}, \bibinfo {author} {\bibfnamefont {E.}~\bibnamefont {Tiesinga}},
  \bibinfo {author} {\bibfnamefont {J.~V.}\ \bibnamefont {Porto}},\ and\
  \bibinfo {author} {\bibfnamefont {C.~J.}\ \bibnamefont {Williams}},\
  }\bibfield  {title} {\bibinfo {title} {Effective three-body interactions of
  neutral bosons in optical lattices},\ }\href
  {https://doi.org/10.1088/1367-2630/11/9/093022} {\bibfield  {journal}
  {\bibinfo  {journal} {New J. Phys.}\ }\textbf {\bibinfo {volume} {11}},\
  \bibinfo {pages} {093022} (\bibinfo {year} {2009})}\BibitemShut {NoStop}%
\bibitem [{\citenamefont {Mazza}\ \emph {et~al.}(2010)\citenamefont {Mazza},
  \citenamefont {Rizzi}, \citenamefont {Lewenstein},\ and\ \citenamefont
  {Cirac}}]{Mazza2010}%
  \BibitemOpen
  \bibfield  {author} {\bibinfo {author} {\bibfnamefont {L.}~\bibnamefont
  {Mazza}}, \bibinfo {author} {\bibfnamefont {M.}~\bibnamefont {Rizzi}},
  \bibinfo {author} {\bibfnamefont {M.}~\bibnamefont {Lewenstein}},\ and\
  \bibinfo {author} {\bibfnamefont {J.~I.}\ \bibnamefont {Cirac}},\ }\bibfield
  {title} {\bibinfo {title} {{Emerging bosons with three-body interactions from
  spin-1 atoms in optical lattices}},\ }\href
  {https://doi.org/10.1103/PhysRevA.82.043629} {\bibfield  {journal} {\bibinfo
  {journal} {Phys. Rev. A}\ }\textbf {\bibinfo {volume} {82}},\ \bibinfo
  {pages} {043629} (\bibinfo {year} {2010})}\BibitemShut {NoStop}%
\bibitem [{\citenamefont {Huerta}\ \emph {et~al.}(2020)\citenamefont {Huerta},
  \citenamefont {Bienias}, \citenamefont {Craddock}, \citenamefont {Gullans},
  \citenamefont {Hachtel}, \citenamefont {Kalinowski}, \citenamefont {Lyon},
  \citenamefont {Gorshkov}, \citenamefont {Rolston},\ and\ \citenamefont
  {Porto}}]{Ornelas-Huerta2020b}%
  \BibitemOpen
  \bibfield  {author} {\bibinfo {author} {\bibfnamefont {D.~P.~O.}\
  \bibnamefont {Huerta}}, \bibinfo {author} {\bibfnamefont {P.}~\bibnamefont
  {Bienias}}, \bibinfo {author} {\bibfnamefont {A.~N.}\ \bibnamefont
  {Craddock}}, \bibinfo {author} {\bibfnamefont {M.~J.}\ \bibnamefont
  {Gullans}}, \bibinfo {author} {\bibfnamefont {A.~J.}\ \bibnamefont
  {Hachtel}}, \bibinfo {author} {\bibfnamefont {M.}~\bibnamefont {Kalinowski}},
  \bibinfo {author} {\bibfnamefont {M.~E.}\ \bibnamefont {Lyon}}, \bibinfo
  {author} {\bibfnamefont {A.~V.}\ \bibnamefont {Gorshkov}}, \bibinfo {author}
  {\bibfnamefont {S.~L.}\ \bibnamefont {Rolston}},\ and\ \bibinfo {author}
  {\bibfnamefont {J.~V.}\ \bibnamefont {Porto}},\ }\bibfield  {title} {\bibinfo
  {title} {Tunable three-body loss in a nonlinear {R}ydberg medium},\ }\href
  {https://arxiv.org/abs/2009.13599} {\bibfield  {journal} {\bibinfo  {journal}
  {arXiv:2009.13599}\ } (\bibinfo {year} {2020})}\BibitemShut {NoStop}%
\bibitem [{\citenamefont {Gorshkov}\ \emph {et~al.}(2013)\citenamefont
  {Gorshkov}, \citenamefont {Nath},\ and\ \citenamefont {Pohl}}]{Gorshkov2013}%
  \BibitemOpen
  \bibfield  {author} {\bibinfo {author} {\bibfnamefont {A.~V.}\ \bibnamefont
  {Gorshkov}}, \bibinfo {author} {\bibfnamefont {R.}~\bibnamefont {Nath}},\
  and\ \bibinfo {author} {\bibfnamefont {T.}~\bibnamefont {Pohl}},\ }\bibfield
  {title} {\bibinfo {title} {{Dissipative many-body quantum optics in Rydberg
  media}},\ }\href {https://doi.org/10.1103/PhysRevLett.110.153601} {\bibfield
  {journal} {\bibinfo  {journal} {Phys. Rev. Lett.}\ }\textbf {\bibinfo
  {volume} {110}},\ \bibinfo {pages} {153601} (\bibinfo {year}
  {2013})}\BibitemShut {NoStop}%
\bibitem [{\citenamefont {Tresp}\ \emph {et~al.}(2015)\citenamefont {Tresp},
  \citenamefont {Bienias}, \citenamefont {Weber}, \citenamefont {Gorniaczyk},
  \citenamefont {Mirgorodskiy}, \citenamefont {B{\"{u}}chler},\ and\
  \citenamefont {Hofferberth}}]{Tresp2015}%
  \BibitemOpen
  \bibfield  {author} {\bibinfo {author} {\bibfnamefont {C.}~\bibnamefont
  {Tresp}}, \bibinfo {author} {\bibfnamefont {P.}~\bibnamefont {Bienias}},
  \bibinfo {author} {\bibfnamefont {S.}~\bibnamefont {Weber}}, \bibinfo
  {author} {\bibfnamefont {H.}~\bibnamefont {Gorniaczyk}}, \bibinfo {author}
  {\bibfnamefont {I.}~\bibnamefont {Mirgorodskiy}}, \bibinfo {author}
  {\bibfnamefont {H.~P.}\ \bibnamefont {B{\"{u}}chler}},\ and\ \bibinfo
  {author} {\bibfnamefont {S.}~\bibnamefont {Hofferberth}},\ }\bibfield
  {title} {\bibinfo {title} {Dipolar dephasing of {R}ydberg d-state
  polaritons},\ }\href {https://doi.org/10.1103/PhysRevLett.115.083602}
  {\bibfield  {journal} {\bibinfo  {journal} {Phys. Rev. Lett.}\ }\textbf
  {\bibinfo {volume} {115}},\ \bibinfo {pages} {083602} (\bibinfo {year}
  {2015})}\BibitemShut {NoStop}%
\bibitem [{\citenamefont {Zeuthen}\ \emph {et~al.}(2017)\citenamefont
  {Zeuthen}, \citenamefont {Gullans}, \citenamefont {Maghrebi},\ and\
  \citenamefont {Gorshkov}}]{Zeuthen2017}%
  \BibitemOpen
  \bibfield  {author} {\bibinfo {author} {\bibfnamefont {E.}~\bibnamefont
  {Zeuthen}}, \bibinfo {author} {\bibfnamefont {M.~J.}\ \bibnamefont
  {Gullans}}, \bibinfo {author} {\bibfnamefont {M.~F.}\ \bibnamefont
  {Maghrebi}},\ and\ \bibinfo {author} {\bibfnamefont {A.~V.}\ \bibnamefont
  {Gorshkov}},\ }\bibfield  {title} {\bibinfo {title} {Correlated photon
  dynamics in dissipative {R}ydberg media},\ }\href
  {https://doi.org/10.1103/PhysRevLett.119.043602} {\bibfield  {journal}
  {\bibinfo  {journal} {Phys. Rev. Lett.}\ }\textbf {\bibinfo {volume} {119}},\
  \bibinfo {pages} {043602} (\bibinfo {year} {2017})}\BibitemShut {NoStop}%
\bibitem [{\citenamefont {Bienias}\ \emph
  {et~al.}(2020{\natexlab{b}})\citenamefont {Bienias}, \citenamefont {Douglas},
  \citenamefont {Paris-Mandoki}, \citenamefont {Titum}, \citenamefont
  {Mirgorodskiy}, \citenamefont {Tresp}, \citenamefont {Zeuthen}, \citenamefont
  {Gullans}, \citenamefont {Manzoni}, \citenamefont {Hofferberth},
  \citenamefont {Chang},\ and\ \citenamefont {Gorshkov}}]{Bienias2020c}%
  \BibitemOpen
  \bibfield  {author} {\bibinfo {author} {\bibfnamefont {P.}~\bibnamefont
  {Bienias}}, \bibinfo {author} {\bibfnamefont {J.}~\bibnamefont {Douglas}},
  \bibinfo {author} {\bibfnamefont {A.}~\bibnamefont {Paris-Mandoki}}, \bibinfo
  {author} {\bibfnamefont {P.}~\bibnamefont {Titum}}, \bibinfo {author}
  {\bibfnamefont {I.}~\bibnamefont {Mirgorodskiy}}, \bibinfo {author}
  {\bibfnamefont {C.}~\bibnamefont {Tresp}}, \bibinfo {author} {\bibfnamefont
  {E.}~\bibnamefont {Zeuthen}}, \bibinfo {author} {\bibfnamefont {M.~J.}\
  \bibnamefont {Gullans}}, \bibinfo {author} {\bibfnamefont {M.}~\bibnamefont
  {Manzoni}}, \bibinfo {author} {\bibfnamefont {S.}~\bibnamefont
  {Hofferberth}}, \bibinfo {author} {\bibfnamefont {D.}~\bibnamefont {Chang}},\
  and\ \bibinfo {author} {\bibfnamefont {A.~V.}\ \bibnamefont {Gorshkov}},\
  }\bibfield  {title} {\bibinfo {title} {{Photon propagation through
  dissipative Rydberg media at large input rates}},\ }\href
  {https://doi.org/10.1103/PhysRevResearch.2.033049} {\bibfield  {journal}
  {\bibinfo  {journal} {Phys. Rev. Res.}\ }\textbf {\bibinfo {volume} {2}},\
  \bibinfo {pages} {033049} (\bibinfo {year} {2020}{\natexlab{b}})}\BibitemShut
  {NoStop}%
\bibitem [{\citenamefont {Roncaglia}\ \emph {et~al.}(2010)\citenamefont
  {Roncaglia}, \citenamefont {Rizzi},\ and\ \citenamefont
  {Cirac}}]{Roncaglia2010}%
  \BibitemOpen
  \bibfield  {author} {\bibinfo {author} {\bibfnamefont {M.}~\bibnamefont
  {Roncaglia}}, \bibinfo {author} {\bibfnamefont {M.}~\bibnamefont {Rizzi}},\
  and\ \bibinfo {author} {\bibfnamefont {J.~I.}\ \bibnamefont {Cirac}},\
  }\bibfield  {title} {\bibinfo {title} {{Pfaffian state generation by strong
  three-body dissipation}},\ }\href
  {https://doi.org/10.1103/PhysRevLett.104.096803} {\bibfield  {journal}
  {\bibinfo  {journal} {Phys. Rev. Lett.}\ }\textbf {\bibinfo {volume} {104}},\
  \bibinfo {pages} {096803} (\bibinfo {year} {2010})}\BibitemShut {NoStop}%
\bibitem [{\citenamefont {Lukin}\ \emph {et~al.}(2001)\citenamefont {Lukin},
  \citenamefont {Fleischhauer}, \citenamefont {Cote}, \citenamefont {Duan},
  \citenamefont {Jaksch}, \citenamefont {Cirac},\ and\ \citenamefont
  {Zoller}}]{Lukin01}%
  \BibitemOpen
  \bibfield  {author} {\bibinfo {author} {\bibfnamefont {M.~D.}\ \bibnamefont
  {Lukin}}, \bibinfo {author} {\bibfnamefont {M.}~\bibnamefont {Fleischhauer}},
  \bibinfo {author} {\bibfnamefont {R.}~\bibnamefont {Cote}}, \bibinfo {author}
  {\bibfnamefont {L.~M.}\ \bibnamefont {Duan}}, \bibinfo {author}
  {\bibfnamefont {D.}~\bibnamefont {Jaksch}}, \bibinfo {author} {\bibfnamefont
  {J.~I.}\ \bibnamefont {Cirac}},\ and\ \bibinfo {author} {\bibfnamefont
  {P.}~\bibnamefont {Zoller}},\ }\bibfield  {title} {\bibinfo {title} {Dipole
  blockade and quantum information processing in mesoscopic atomic ensembles},\
  }\href {https://doi.org/10.1103/PhysRevLett.87.037901} {\bibfield  {journal}
  {\bibinfo  {journal} {Phys. Rev. Lett.}\ }\textbf {\bibinfo {volume} {87}},\
  \bibinfo {pages} {037901} (\bibinfo {year} {2001})}\BibitemShut {NoStop}%
\bibitem [{\citenamefont {Lewenstein}\ \emph {et~al.}(2007)\citenamefont
  {Lewenstein}, \citenamefont {Sanpera}, \citenamefont {Ahufinger},
  \citenamefont {Damski}, \citenamefont {Sen},\ and\ \citenamefont
  {Sen}}]{Lewenstein2007}%
  \BibitemOpen
  \bibfield  {author} {\bibinfo {author} {\bibfnamefont {M.}~\bibnamefont
  {Lewenstein}}, \bibinfo {author} {\bibfnamefont {A.}~\bibnamefont {Sanpera}},
  \bibinfo {author} {\bibfnamefont {V.}~\bibnamefont {Ahufinger}}, \bibinfo
  {author} {\bibfnamefont {B.}~\bibnamefont {Damski}}, \bibinfo {author}
  {\bibfnamefont {A.}~\bibnamefont {Sen}},\ and\ \bibinfo {author}
  {\bibfnamefont {U.}~\bibnamefont {Sen}},\ }\bibfield  {title} {\bibinfo
  {title} {{Ultracold atomic gases in optical lattices: Mimicking condensed
  matter physics and beyond}},\ }\href
  {https://doi.org/10.1080/00018730701223200} {\bibfield  {journal} {\bibinfo
  {journal} {Adv. Phys.}\ }\textbf {\bibinfo {volume} {56}},\ \bibinfo {pages}
  {243} (\bibinfo {year} {2007})}\BibitemShut {NoStop}%
\bibitem [{\citenamefont {Kalinowski}\ \emph {et~al.}()\citenamefont
  {Kalinowski} \emph {et~al.}}]{Kalinowski2020}%
  \BibitemOpen
  \bibfield  {author} {\bibinfo {author} {\bibfnamefont {M.}~\bibnamefont
  {Kalinowski}} \emph {et~al.},\ }\href@noop {} {\bibinfo  {journal} {(in
  preparation)}\ }\BibitemShut {NoStop}%
\bibitem [{\citenamefont {Sommer}\ \emph {et~al.}()\citenamefont {Sommer},
  \citenamefont {B{\"{u}}chler},\ and\ \citenamefont {Simon}}]{Sommer2015}%
  \BibitemOpen
\bibfield  {journal} {  }\bibfield  {author} {\bibinfo {author} {\bibfnamefont
  {A.}~\bibnamefont {Sommer}}, \bibinfo {author} {\bibfnamefont {H.~P.}\
  \bibnamefont {B{\"{u}}chler}},\ and\ \bibinfo {author} {\bibfnamefont
  {J.}~\bibnamefont {Simon}},\ }\bibfield  {title} {\bibinfo {title} {Quantum
  crystals and {L}aughlin droplets of cavity {R}ydberg polaritons},\ }\Eprint
  {https://arxiv.org/abs/1506.00341} {arXiv:1506.00341 [cond-mat.quant-gas]}
  \BibitemShut {NoStop}%
\bibitem [{\citenamefont {Parigi}\ \emph {et~al.}(2012)\citenamefont {Parigi},
  \citenamefont {Bimbard}, \citenamefont {Stanojevic}, \citenamefont
  {Hilliard}, \citenamefont {Nogrette}, \citenamefont {Tualle-Brouri},
  \citenamefont {Ourjoumtsev},\ and\ \citenamefont {Grangier}}]{Parigi2012}%
  \BibitemOpen
  \bibfield  {author} {\bibinfo {author} {\bibfnamefont {V.}~\bibnamefont
  {Parigi}}, \bibinfo {author} {\bibfnamefont {E.}~\bibnamefont {Bimbard}},
  \bibinfo {author} {\bibfnamefont {J.}~\bibnamefont {Stanojevic}}, \bibinfo
  {author} {\bibfnamefont {A.~J.}\ \bibnamefont {Hilliard}}, \bibinfo {author}
  {\bibfnamefont {F.}~\bibnamefont {Nogrette}}, \bibinfo {author}
  {\bibfnamefont {R.}~\bibnamefont {Tualle-Brouri}}, \bibinfo {author}
  {\bibfnamefont {A.}~\bibnamefont {Ourjoumtsev}},\ and\ \bibinfo {author}
  {\bibfnamefont {P.}~\bibnamefont {Grangier}},\ }\bibfield  {title} {\bibinfo
  {title} {{Observation and measurement of "giant" dispersive optical
  non-linearities in an ensemble of cold Rydberg atoms}},\ }\href
  {https://doi.org/10.1103/PhysRevLett.109.233602} {\bibfield  {journal}
  {\bibinfo  {journal} {Phys. Rev. Lett}\ }\textbf {\bibinfo {volume} {109}},\
  \bibinfo {pages} {233602} (\bibinfo {year} {2012})}\BibitemShut {NoStop}%
\bibitem [{\citenamefont {Georgakopoulos}\ \emph {et~al.}(2018)\citenamefont
  {Georgakopoulos}, \citenamefont {Sommer},\ and\ \citenamefont
  {Simon}}]{Georgakopoulos2018}%
  \BibitemOpen
  \bibfield  {author} {\bibinfo {author} {\bibfnamefont {A.}~\bibnamefont
  {Georgakopoulos}}, \bibinfo {author} {\bibfnamefont {A.}~\bibnamefont
  {Sommer}},\ and\ \bibinfo {author} {\bibfnamefont {J.}~\bibnamefont
  {Simon}},\ }\bibfield  {title} {\bibinfo {title} {Theory of interacting
  cavity {R}ydberg polaritons},\ }\href
  {https://doi.org/10.1088/2058-9565/aadf4d} {\bibfield  {journal} {\bibinfo
  {journal} {Quantum Sci. Technol.}\ }\textbf {\bibinfo {volume} {4}},\
  \bibinfo {pages} {014005} (\bibinfo {year} {2018})}\BibitemShut {NoStop}%
\bibitem [{\citenamefont {Litinskaya}\ \emph {et~al.}(2016)\citenamefont
  {Litinskaya}, \citenamefont {Tignone},\ and\ \citenamefont
  {Pupillo}}]{Litinskaya2016}%
  \BibitemOpen
  \bibfield  {author} {\bibinfo {author} {\bibfnamefont {M.}~\bibnamefont
  {Litinskaya}}, \bibinfo {author} {\bibfnamefont {E.}~\bibnamefont
  {Tignone}},\ and\ \bibinfo {author} {\bibfnamefont {G.}~\bibnamefont
  {Pupillo}},\ }\bibfield  {title} {\bibinfo {title} {{Cavity polaritons with
  Rydberg blockade and long-range interactions}},\ }\href
  {https://doi.org/10.1088/0953-4075/49/16/164006} {\bibfield  {journal}
  {\bibinfo  {journal} {J. Phys. B}\ }\textbf {\bibinfo {volume} {49}},\
  \bibinfo {pages} {164006} (\bibinfo {year} {2016})}\BibitemShut {NoStop}%
\bibitem [{\citenamefont {Grankin}\ \emph {et~al.}(2014)\citenamefont
  {Grankin}, \citenamefont {Brion}, \citenamefont {Bimbard}, \citenamefont
  {Boddeda}, \citenamefont {Usmani}, \citenamefont {Ourjoumtsev},\ and\
  \citenamefont {Grangier}}]{Grankin2014}%
  \BibitemOpen
  \bibfield  {author} {\bibinfo {author} {\bibfnamefont {A.}~\bibnamefont
  {Grankin}}, \bibinfo {author} {\bibfnamefont {E.}~\bibnamefont {Brion}},
  \bibinfo {author} {\bibfnamefont {E.}~\bibnamefont {Bimbard}}, \bibinfo
  {author} {\bibfnamefont {R.}~\bibnamefont {Boddeda}}, \bibinfo {author}
  {\bibfnamefont {I.}~\bibnamefont {Usmani}}, \bibinfo {author} {\bibfnamefont
  {A.}~\bibnamefont {Ourjoumtsev}},\ and\ \bibinfo {author} {\bibfnamefont
  {P.}~\bibnamefont {Grangier}},\ }\bibfield  {title} {\bibinfo {title}
  {{Quantum statistics of light transmitted through an intracavity Rydberg
  medium}},\ }\href {https://doi.org/10.1088/1367-2630/16/4/043020} {\bibfield
  {journal} {\bibinfo  {journal} {New J. Phys.}\ }\textbf {\bibinfo {volume}
  {16}},\ \bibinfo {pages} {043020} (\bibinfo {year} {2014})}\BibitemShut
  {NoStop}%
\bibitem [{\citenamefont {Bienias}\ \emph {et~al.}(2014)\citenamefont
  {Bienias}, \citenamefont {Choi}, \citenamefont {Firstenberg}, \citenamefont
  {Maghrebi}, \citenamefont {Gullans}, \citenamefont {Lukin}, \citenamefont
  {Gorshkov},\ and\ \citenamefont {B{\"{u}}chler}}]{Bienias2014}%
  \BibitemOpen
  \bibfield  {author} {\bibinfo {author} {\bibfnamefont {P.}~\bibnamefont
  {Bienias}}, \bibinfo {author} {\bibfnamefont {S.}~\bibnamefont {Choi}},
  \bibinfo {author} {\bibfnamefont {O.}~\bibnamefont {Firstenberg}}, \bibinfo
  {author} {\bibfnamefont {M.~F.}\ \bibnamefont {Maghrebi}}, \bibinfo {author}
  {\bibfnamefont {M.}~\bibnamefont {Gullans}}, \bibinfo {author} {\bibfnamefont
  {M.~D.}\ \bibnamefont {Lukin}}, \bibinfo {author} {\bibfnamefont {A.~V.}\
  \bibnamefont {Gorshkov}},\ and\ \bibinfo {author} {\bibfnamefont {H.~P.}\
  \bibnamefont {B{\"{u}}chler}},\ }\bibfield  {title} {\bibinfo {title}
  {{Scattering resonances and bound states for strongly interacting Rydberg
  polaritons}},\ }\href {https://doi.org/10.1103/PhysRevA.90.053804} {\bibfield
   {journal} {\bibinfo  {journal} {Phys. Rev. A}\ }\textbf {\bibinfo {volume}
  {90}},\ \bibinfo {pages} {053804} (\bibinfo {year} {2014})}\BibitemShut
  {NoStop}%
\bibitem [{\citenamefont {Gorshkov}\ \emph {et~al.}(2011)\citenamefont
  {Gorshkov}, \citenamefont {Otterbach}, \citenamefont {Fleischhauer},
  \citenamefont {Pohl},\ and\ \citenamefont {Lukin}}]{Gorshkov2011}%
  \BibitemOpen
  \bibfield  {author} {\bibinfo {author} {\bibfnamefont {A.~V.}\ \bibnamefont
  {Gorshkov}}, \bibinfo {author} {\bibfnamefont {J.}~\bibnamefont {Otterbach}},
  \bibinfo {author} {\bibfnamefont {M.}~\bibnamefont {Fleischhauer}}, \bibinfo
  {author} {\bibfnamefont {T.}~\bibnamefont {Pohl}},\ and\ \bibinfo {author}
  {\bibfnamefont {M.~D.}\ \bibnamefont {Lukin}},\ }\bibfield  {title} {\bibinfo
  {title} {{Photon-photon interactions via Rydberg blockade}},\ }\href
  {https://doi.org/10.1103/PhysRevLett.107.133602} {\bibfield  {journal}
  {\bibinfo  {journal} {Phys. Rev. Lett.}\ }\textbf {\bibinfo {volume} {107}},\
  \bibinfo {pages} {133602} (\bibinfo {year} {2011})}\BibitemShut {NoStop}%
\bibitem [{\citenamefont {Bienias}\ and\ \citenamefont
  {B{\"{u}}chler}(2020)}]{Bienias2020}%
  \BibitemOpen
  \bibfield  {author} {\bibinfo {author} {\bibfnamefont {P.}~\bibnamefont
  {Bienias}}\ and\ \bibinfo {author} {\bibfnamefont {H.~P.}\ \bibnamefont
  {B{\"{u}}chler}},\ }\bibfield  {title} {\bibinfo {title} {{Two photon
  conditional phase gate based on Rydberg slow light polaritons}},\ }\href
  {https://doi.org/10.1088/1361-6455/ab5bed} {\bibfield  {journal} {\bibinfo
  {journal} {J. Phys. B At. Mol. Opt. Phys.}\ }\textbf {\bibinfo {volume}
  {53}},\ \bibinfo {pages} {54003} (\bibinfo {year} {2020})}\BibitemShut
  {NoStop}%
\bibitem [{\citenamefont {Carmichael}(1993)}]{Carmichael1993}%
  \BibitemOpen
  \bibfield  {author} {\bibinfo {author} {\bibfnamefont {H.}~\bibnamefont
  {Carmichael}},\ }\href@noop {} {\emph {\bibinfo {title} {{An open systems
  approach to quantum optics: lectures presented at the Universit{\'e} libre de
  Bruxelles, October 28 to November 4, 1991}}}}\ (\bibinfo  {publisher}
  {Springer-Verlag},\ \bibinfo {year} {1993})\BibitemShut {NoStop}%
\bibitem [{sup()}]{supplement}%
  \BibitemOpen
  \href@noop {} {}\bibinfo {howpublished} {See Supplemental Material at [URL
  will be inserted by publisher] for detailed derivations of energy shifts and
  transmission properties.}\BibitemShut {Stop}%
\bibitem [{\citenamefont {Bienias}\ and\ \citenamefont
  {B{\"{u}}chler}(2016)}]{Bienias2016}%
  \BibitemOpen
  \bibfield  {author} {\bibinfo {author} {\bibfnamefont {P.}~\bibnamefont
  {Bienias}}\ and\ \bibinfo {author} {\bibfnamefont {H.~P.}\ \bibnamefont
  {B{\"{u}}chler}},\ }\bibfield  {title} {\bibinfo {title} {{Quantum theory of
  Kerr nonlinearity with Rydberg slow light polaritons}},\ }\href
  {https://doi.org/10.1088/1367-2630/aa50c3} {\bibfield  {journal} {\bibinfo
  {journal} {New J. Phys.}\ }\textbf {\bibinfo {volume} {18}},\ \bibinfo
  {pages} {123026} (\bibinfo {year} {2016})}\BibitemShut {NoStop}%
\bibitem [{\citenamefont {Bienias}(2016)}]{Bienias2016a}%
  \BibitemOpen
  \bibfield  {author} {\bibinfo {author} {\bibfnamefont {P.}~\bibnamefont
  {Bienias}},\ }\bibfield  {title} {\bibinfo {title} {{Few-body quantum physics
  with strongly interacting Rydberg polaritons}},\ }\href
  {https://doi.org/10.1140/epjst/e2016-60004-x} {\bibfield  {journal} {\bibinfo
   {journal} {Eur. Phys. J. Spec. Top.}\ }\textbf {\bibinfo {volume} {225}},\
  \bibinfo {pages} {2957} (\bibinfo {year} {2016})}\BibitemShut {NoStop}%
\bibitem [{\citenamefont {Faddeev}(1961)}]{Faddeev1961}%
  \BibitemOpen
  \bibfield  {author} {\bibinfo {author} {\bibfnamefont {L.~D.}\ \bibnamefont
  {Faddeev}},\ }\bibfield  {title} {\bibinfo {title} {{Scattering theory for a
  three-particle system}},\ }\href {https://doi.org/10.1142/9789814340960_0004}
  {\bibfield  {journal} {\bibinfo  {journal} {Sov. Phys. JETP}\ }\textbf
  {\bibinfo {volume} {12}},\ \bibinfo {pages} {1014} (\bibinfo {year}
  {1961})}\BibitemShut {NoStop}%
\bibitem [{\citenamefont {Wang}\ \emph {et~al.}()\citenamefont {Wang} \emph
  {et~al.}}]{Wang2020}%
  \BibitemOpen
  \bibfield  {author} {\bibinfo {author} {\bibfnamefont {Y.}~\bibnamefont
  {Wang}} \emph {et~al.},\ }\href@noop {} {\bibinfo  {journal} {(in
  preparation)}\ }\BibitemShut {NoStop}%
\bibitem [{\citenamefont {Murray}\ and\ \citenamefont
  {Pohl}(2016)}]{Murray2016}%
  \BibitemOpen
\bibfield  {journal} {  }\bibfield  {author} {\bibinfo {author} {\bibfnamefont
  {C.}~\bibnamefont {Murray}}\ and\ \bibinfo {author} {\bibfnamefont
  {T.}~\bibnamefont {Pohl}},\ }\bibfield  {title} {\bibinfo {title} {Chapter
  seven - quantum and nonlinear optics in strongly interacting atomic
  ensembles}\ }(\bibinfo  {publisher} {Academic Press},\ \bibinfo {year}
  {2016})\ pp.\ \bibinfo {pages} {321 -- 372}\BibitemShut {NoStop}%
\bibitem [{\citenamefont {Manzoni}\ \emph {et~al.}(2017)\citenamefont
  {Manzoni}, \citenamefont {Chang},\ and\ \citenamefont
  {Douglas}}]{Manzoni2017}%
  \BibitemOpen
  \bibfield  {author} {\bibinfo {author} {\bibfnamefont {M.~T.}\ \bibnamefont
  {Manzoni}}, \bibinfo {author} {\bibfnamefont {D.~E.}\ \bibnamefont {Chang}},\
  and\ \bibinfo {author} {\bibfnamefont {J.~S.}\ \bibnamefont {Douglas}},\
  }\bibfield  {title} {\bibinfo {title} {{Simulating quantum light propagation
  through atomic ensembles using matrix product states}},\ }\href
  {https://doi.org/10.1038/s41467-017-01416-4} {\bibfield  {journal} {\bibinfo
  {journal} {Nat. Commun.}\ }\textbf {\bibinfo {volume} {8}},\ \bibinfo {pages}
  {1743} (\bibinfo {year} {2017})}\BibitemShut {NoStop}%
\bibitem [{\citenamefont {Grass}\ \emph {et~al.}(2018)\citenamefont {Grass},
  \citenamefont {Bienias}, \citenamefont {Gullans}, \citenamefont {Lundgren},
  \citenamefont {Maciejko},\ and\ \citenamefont {Gorshkov}}]{Grass2018}%
  \BibitemOpen
  \bibfield  {author} {\bibinfo {author} {\bibfnamefont {T.}~\bibnamefont
  {Grass}}, \bibinfo {author} {\bibfnamefont {P.}~\bibnamefont {Bienias}},
  \bibinfo {author} {\bibfnamefont {M.~J.}\ \bibnamefont {Gullans}}, \bibinfo
  {author} {\bibfnamefont {R.}~\bibnamefont {Lundgren}}, \bibinfo {author}
  {\bibfnamefont {J.}~\bibnamefont {Maciejko}},\ and\ \bibinfo {author}
  {\bibfnamefont {A.~V.}\ \bibnamefont {Gorshkov}},\ }\bibfield  {title}
  {\bibinfo {title} {Fractional quantum {H}all phases of bosons with tunable
  interactions: From the {L}aughlin liquid to a fractional {W}igner crystal},\
  }\href {https://doi.org/10.1103/PhysRevLett.121.253403} {\bibfield  {journal}
  {\bibinfo  {journal} {Phys. Rev. Lett.}\ }\textbf {\bibinfo {volume} {121}},\
  \bibinfo {pages} {253403} (\bibinfo {year} {2018})}\BibitemShut {NoStop}%
\end{thebibliography}%


%Control: production of eprint (1) enabled
\begin{thebibliography}{7}%
\makeatletter
\providecommand \@ifxundefined [1]{%
 \@ifx{#1\undefined}
}%
\providecommand \@ifnum [1]{%
 \ifnum #1\expandafter \@firstoftwo
 \else \expandafter \@secondoftwo
 \fi
}%
\providecommand \@ifx [1]{%
 \ifx #1\expandafter \@firstoftwo
 \else \expandafter \@secondoftwo
 \fi
}%
\providecommand \natexlab [1]{#1}%
\providecommand \enquote  [1]{``#1''}%
\providecommand \bibnamefont  [1]{#1}%
\providecommand \bibfnamefont [1]{#1}%
\providecommand \citenamefont [1]{#1}%
\providecommand \href@noop [0]{\@secondoftwo}%
\providecommand \href [0]{\begingroup \@sanitize@url \@href}%
\providecommand \@href[1]{\@@startlink{#1}\@@href}%
\providecommand \@@href[1]{\endgroup#1\@@endlink}%
\providecommand \@sanitize@url [0]{\catcode `\\12\catcode `\$12\catcode
  `\&12\catcode `\#12\catcode `\^12\catcode `\_12\catcode `\%12\relax}%
\providecommand \@@startlink[1]{}%
\providecommand \@@endlink[0]{}%
\providecommand \url  [0]{\begingroup\@sanitize@url \@url }%
\providecommand \@url [1]{\endgroup\@href {#1}{\urlprefix }}%
\providecommand \urlprefix  [0]{URL }%
\providecommand \Eprint [0]{\href }%
\providecommand \doibase [0]{https://doi.org/}%
\providecommand \selectlanguage [0]{\@gobble}%
\providecommand \bibinfo  [0]{\@secondoftwo}%
\providecommand \bibfield  [0]{\@secondoftwo}%
\providecommand \translation [1]{[#1]}%
\providecommand \BibitemOpen [0]{}%
\providecommand \bibitemStop [0]{}%
\providecommand \bibitemNoStop [0]{.\EOS\space}%
\providecommand \EOS [0]{\spacefactor3000\relax}%
\providecommand \BibitemShut  [1]{\csname bibitem#1\endcsname}%
\let\auto@bib@innerbib\@empty
%</preamble>
\bibitem [{\citenamefont {Bienias}\ \emph {et~al.}(2014)\citenamefont
  {Bienias}, \citenamefont {Choi}, \citenamefont {Firstenberg}, \citenamefont
  {Maghrebi}, \citenamefont {Gullans}, \citenamefont {Lukin}, \citenamefont
  {Gorshkov},\ and\ \citenamefont {B{\"{u}}chler}}]{Bienias2014}%
  \BibitemOpen
  \bibfield  {author} {\bibinfo {author} {\bibfnamefont {P.}~\bibnamefont
  {Bienias}}, \bibinfo {author} {\bibfnamefont {S.}~\bibnamefont {Choi}},
  \bibinfo {author} {\bibfnamefont {O.}~\bibnamefont {Firstenberg}}, \bibinfo
  {author} {\bibfnamefont {M.~F.}\ \bibnamefont {Maghrebi}}, \bibinfo {author}
  {\bibfnamefont {M.}~\bibnamefont {Gullans}}, \bibinfo {author} {\bibfnamefont
  {M.~D.}\ \bibnamefont {Lukin}}, \bibinfo {author} {\bibfnamefont {A.~V.}\
  \bibnamefont {Gorshkov}},\ and\ \bibinfo {author} {\bibfnamefont {H.~P.}\
  \bibnamefont {B{\"{u}}chler}},\ }\bibfield  {title} {\bibinfo {title}
  {{Scattering resonances and bound states for strongly interacting Rydberg
  polaritons}},\ }\href {https://doi.org/10.1103/PhysRevA.90.053804} {\bibfield
   {journal} {\bibinfo  {journal} {Phys. Rev. A}\ }\textbf {\bibinfo {volume}
  {90}},\ \bibinfo {pages} {053804} (\bibinfo {year} {2014})}\BibitemShut
  {NoStop}%
\bibitem [{\citenamefont {Jachymski}\ \emph {et~al.}(2016)\citenamefont
  {Jachymski}, \citenamefont {Bienias},\ and\ \citenamefont
  {B{\"{u}}chler}}]{Jachymski2016}%
  \BibitemOpen
  \bibfield  {author} {\bibinfo {author} {\bibfnamefont {K.}~\bibnamefont
  {Jachymski}}, \bibinfo {author} {\bibfnamefont {P.}~\bibnamefont {Bienias}},\
  and\ \bibinfo {author} {\bibfnamefont {H.~P.}\ \bibnamefont
  {B{\"{u}}chler}},\ }\bibfield  {title} {\bibinfo {title} {{Three-body
  interactions of slow light Rydberg polaritons}},\ }\href
  {https://doi.org/10.1103/PhysRevLett.117.053601} {\bibfield  {journal}
  {\bibinfo  {journal} {Phys. Rev. Lett.}\ }\textbf {\bibinfo {volume} {117}},\
  \bibinfo {pages} {053601} (\bibinfo {year} {2016})}\BibitemShut {NoStop}%
\bibitem [{\citenamefont {Gullans}\ \emph {et~al.}(2016)\citenamefont
  {Gullans}, \citenamefont {Thompson}, \citenamefont {Wang}, \citenamefont
  {Liang}, \citenamefont {Vuleti{\'{c}}}, \citenamefont {Lukin},\ and\
  \citenamefont {Gorshkov}}]{Gullans2016}%
  \BibitemOpen
  \bibfield  {author} {\bibinfo {author} {\bibfnamefont {M.~J.}\ \bibnamefont
  {Gullans}}, \bibinfo {author} {\bibfnamefont {J.~D.}\ \bibnamefont
  {Thompson}}, \bibinfo {author} {\bibfnamefont {Y.}~\bibnamefont {Wang}},
  \bibinfo {author} {\bibfnamefont {Q.~Y.}\ \bibnamefont {Liang}}, \bibinfo
  {author} {\bibfnamefont {V.}~\bibnamefont {Vuleti{\'{c}}}}, \bibinfo {author}
  {\bibfnamefont {M.~D.}\ \bibnamefont {Lukin}},\ and\ \bibinfo {author}
  {\bibfnamefont {A.~V.}\ \bibnamefont {Gorshkov}},\ }\bibfield  {title}
  {\bibinfo {title} {{Effective Field Theory for Rydberg Polaritons}},\ }\href
  {https://doi.org/10.1103/PhysRevLett.117.113601} {\bibfield  {journal}
  {\bibinfo  {journal} {Phys. Rev. Lett.}\ }\textbf {\bibinfo {volume} {117}},\
  \bibinfo {pages} {113601} (\bibinfo {year} {2016})}\BibitemShut {NoStop}%
\bibitem [{\citenamefont {Kalinowski}\ \emph {et~al.}()\citenamefont
  {Kalinowski} \emph {et~al.}}]{Kalinowski2020}%
  \BibitemOpen
  \bibfield  {author} {\bibinfo {author} {\bibfnamefont {M.}~\bibnamefont
  {Kalinowski}} \emph {et~al.},\ }\href@noop {} {\bibinfo  {journal} {(in
  preparation)}\ }\BibitemShut {NoStop}%
\bibitem [{\citenamefont {Liang}\ \emph {et~al.}(2018)\citenamefont {Liang},
  \citenamefont {Venkatramani}, \citenamefont {Cantu}, \citenamefont
  {Nicholson}, \citenamefont {Gullans}, \citenamefont {Gorshkov}, \citenamefont
  {Thompson}, \citenamefont {Chin}, \citenamefont {Lukin},\ and\ \citenamefont
  {Vuleti{\'{c}}}}]{Liang2018}%
  \BibitemOpen
\bibfield  {journal} {  }\bibfield  {author} {\bibinfo {author} {\bibfnamefont
  {Q.-Y.}\ \bibnamefont {Liang}}, \bibinfo {author} {\bibfnamefont {A.~V.}\
  \bibnamefont {Venkatramani}}, \bibinfo {author} {\bibfnamefont {S.~H.}\
  \bibnamefont {Cantu}}, \bibinfo {author} {\bibfnamefont {T.~L.}\ \bibnamefont
  {Nicholson}}, \bibinfo {author} {\bibfnamefont {M.~J.}\ \bibnamefont
  {Gullans}}, \bibinfo {author} {\bibfnamefont {A.~V.}\ \bibnamefont
  {Gorshkov}}, \bibinfo {author} {\bibfnamefont {J.~D.}\ \bibnamefont
  {Thompson}}, \bibinfo {author} {\bibfnamefont {C.}~\bibnamefont {Chin}},
  \bibinfo {author} {\bibfnamefont {M.~D.}\ \bibnamefont {Lukin}},\ and\
  \bibinfo {author} {\bibfnamefont {V.}~\bibnamefont {Vuleti{\'{c}}}},\
  }\bibfield  {title} {\bibinfo {title} {{Observation of three-photon bound
  states in a quantum nonlinear medium}},\ }\href
  {https://doi.org/10.1126/science.aao7293} {\bibfield  {journal} {\bibinfo
  {journal} {Science}\ }\textbf {\bibinfo {volume} {359}},\ \bibinfo {pages}
  {783} (\bibinfo {year} {2018})}\BibitemShut {NoStop}%
\bibitem [{\citenamefont {Faddeev}(1961)}]{Faddeev1961}%
  \BibitemOpen
  \bibfield  {author} {\bibinfo {author} {\bibfnamefont {L.~D.}\ \bibnamefont
  {Faddeev}},\ }\bibfield  {title} {\bibinfo {title} {{Scattering theory for a
  three-particle system}},\ }\href {https://doi.org/10.1142/9789814340960_0004}
  {\bibfield  {journal} {\bibinfo  {journal} {Sov. Phys. JETP}\ }\textbf
  {\bibinfo {volume} {12}},\ \bibinfo {pages} {1014} (\bibinfo {year}
  {1961})}\BibitemShut {NoStop}%
\bibitem [{\citenamefont {Wang}\ \emph {et~al.}()\citenamefont {Wang} \emph
  {et~al.}}]{Wang2020}%
  \BibitemOpen
  \bibfield  {author} {\bibinfo {author} {\bibfnamefont {Y.}~\bibnamefont
  {Wang}} \emph {et~al.},\ }\href@noop {} {\bibinfo  {journal} {(in
  preparation)}\ }\BibitemShut {NoStop}%
\end{thebibliography}%

\ifsupp
\def\thesection{\Roman{section}}
\setcounter{secnumdepth}{2}
\widetext
\pagebreak
\ExecuteMetaData[supplement]{document}
\fi
\end{document}

\renewcommand{\bibnumfmt}[1]{[S#1]}
\renewcommand{\citenumfont}[1]{S#1}
\title{Resonant enhancement of three-body loss between  strongly interacting photons} 
	
	\author{Marcin Kalinowski}
		\thanks{These two authors contributed equally.}
	\affiliation{Joint Quantum Institute, NIST/University of Maryland, College Park, Maryland 20742 USA}
	\affiliation{Faculty of Physics, University of Warsaw, Pasteura 5, 02-093 Warsaw, Poland}
	\author{Yidan Wang}
	\thanks{These two authors contributed equally.}
	\affiliation{Joint Quantum Institute, NIST/University of Maryland, College Park, Maryland 20742 USA}
	 \author{Przemyslaw Bienias}
	 \affiliation{Joint Quantum Institute, NIST/University of Maryland, College Park, Maryland 20742 USA}
	 \affiliation{Joint Center for Quantum Information and Computer Science, NIST/University of Maryland, College Park, Maryland 20742 USA}
	 \author{ Michael J. Gullans}
	 \affiliation{Joint Quantum Institute, NIST/University of Maryland, College Park, Maryland 20742 USA}
	 \affiliation{Joint Center for Quantum Information and Computer Science, NIST/University of Maryland, College Park, Maryland 20742 USA}
	 \affiliation{Department of Physics, Princeton University, Princeton, New Jersey 08544 USA}
	 \author{Dalia P. Ornelas-Huerta}	 
	 \affiliation{Joint Quantum Institute, NIST/University of Maryland, College Park, Maryland 20742 USA}
	 \author{ Alexander N. Craddock}
	 \affiliation{Joint Quantum Institute, NIST/University of Maryland, College Park, Maryland 20742 USA}
	 \author{Steven L. Rolston}
	  \affiliation{Joint Quantum Institute, NIST/University of Maryland, College Park, Maryland 20742 USA}
	 \author{J. V. Porto}
	 \affiliation{Joint Quantum Institute, NIST/University of Maryland, College Park, Maryland 20742 USA}
	 \author{ Hans Peter B{\" u}chler}
	 \affiliation{Institute for Theoretical Physics III and Center for Integrated Quantum Science and Technology, University of Stuttgart, 70550 Stuttgart, Germany}
	 \author{Alexey V. Gorshkov}
	 \affiliation{Joint Quantum Institute, NIST/University of Maryland, College Park, Maryland 20742 USA}
	 \affiliation{Joint Center for Quantum Information and Computer Science, NIST/University of Maryland, College Park, Maryland 20742 USA}
	 \author{}
%<*document>
\makeatletter
\begingroup
\ltx@footnote@pop
\def\@mpfn{mpfootnote}%
\def\thempfn{\thempfootnote}%
\c@mpfootnote\z@
\let\@makefnmark\frontmatter@makefnmark
%\frontmatter@setup
%\thispagestyle{titlepage}\label{FirstPage}%
\begingroup
\frontmatter@title@above
\frontmatter@title@format
Supplemental Material for ``\@title"
\unskip
\phantomsection\expandafter\@argswap@val\expandafter{\@title}{\addcontentsline{toc}{title}}%
\@ifx{\@title@aux\@title@aux@cleared}{}{%
	\expandafter\frontmatter@footnote\expandafter{\@title@aux}%
}%
\par
\frontmatter@title@below
\endgroup
\groupauthors@sw{%
	\frontmatter@author@produce@group
}{%
	\frontmatter@author@produce@script
}%

\par
\frontmatter@finalspace
\endgroup

\makeatother

    \setcounter{equation}{0}
    \setcounter{figure}{0}
    \setcounter{table}{0}
    
    \renewcommand{\theequation}{S\arabic{equation}}
    \renewcommand{\thefigure}{S\arabic{figure}}
    %%%%%%%%%% Prefix a "S" to all equations, figures, tables and reset the counter %%%%%%%%%%
	\par In this Supplemental Material, we provide detailed calculations used to obtain results mentioned in the main text. This covers the in-depth introduction of  the microscopic model and related quantities (Sec.~\ref{Ssec:1}), the solution to the two-body problem (Sec.~\ref{Ssec:2}), the solution of three-body problem (Sec.~\ref{Ssec:3}), and the description of our effective transmission model (Sec.~\ref{Ssec:4}).
	
    \section{Running-wave cavity Hamiltonian}\label{Ssec:1}
    In this section, we introduce the details of the microscopic model used in this work.
    
    \par We consider a running-wave optical cavity in one dimension, supporting a single photonic mode. Moreover, we assume a constant density of the atomic cloud, so that the system is translationally invariant. The cavity mode with the profile $u_0(z)$ is created by the operator $ a^\dagger$. While there is only one photonic mode present, the atomic medium can support a broad range of excitations, which are captured by introducing additional mode functions $ u_{q\neq 0}(z) $.  Together,  $\{u_0,u_{q\neq0}\}$ form an orthonormal basis and can be used to express various field operators
	\begin{align}
	    E^\dagger(z) &= u_0^\ast(z)a^\dagger, & P^\dagger_{q} &= \int dz \,u_{q}(z) P^\dagger(z), & S^\dagger_{q} &= \int dz\, u_q(z) e^{ik_c z} S^\dagger(z),
	\end{align}
	where $ E^\dagger(z) $ creates the cavity photon at position $ z $, while $ P^\dagger(z) $ and $ S^\dagger(z) $ create an excitation of the medium at position $ z $ to the  atomic state $ \ket{p} $ ans $ \ket{s} $, respectively. Thanks to the translational symmetry, we can identify index $ q $ with momentum and write the explicit form of  these mode functions:  $u_q(z)=\frac{1}{\sqrt{L}}e^{i q z}$. This way, the momentum $ q=0 $ corresponds to the cavity photon in the rotating frame. The Hamiltonian for such a system is
	\begin{equation}
	\label{Seq:H1M}
	    H_{\rm cav} = \underbrace{\begin{pmatrix}a\\P_0\\S_0\end{pmatrix}^\dagger
	    \begin{pmatrix}\delta_r&g & 0  \\
	   g&\Delta& \Omega \\
	  0&\Omega& -i\gamma_s
	                   \end{pmatrix}
	    \begin{pmatrix}a\\P_0\\S_0\end{pmatrix}}_{\text{Photon mode}}
	    + \underbrace{\sum_{q\neq0}\begin{pmatrix}P_q\\S_q\end{pmatrix}^\dagger
	    \begin{pmatrix}\Delta&\Omega  \\
	                   \Omega&-i\gamma_s
	                   \end{pmatrix}
	    \begin{pmatrix}P_q\\S_q\end{pmatrix}}_{\text{Spin waves}}+H_{\rm int},
	\end{equation}
	where $\Delta = \delta - i\gamma$ is the complex detuning of the classical field with Rabi frequency $\Omega$, $2\gamma_s$ is the decay rate of the Rydberg state, and $\delta_r$ is the (two-photon) detuning from the EIT resonance. The interaction Hamiltonian $H_{\rm int}$ is 
    \begin{equation}
        H_{\rm int} = \frac{1}{2}\int dz dz' S^\dagger(z)S^\dagger(z')V(z-z')S(z')S(z')   = \frac{1}{2}\sum_{q_1,q_2}S^\dagger_{q_1} S^\dagger_{-q_1} S_{q_2} S_{-q_2} \int \frac{dr}{L} \, V(r)e^{i (q_1-q_2) r},
    \end{equation}
	where second equality holds for  vanishing total momentum $K=0$, which is the case we consider here (as we will explain in Sec.~\ref{Ssec:3}, only the $K=0$ solution to the two-body problem will be necessary for our solution to the three-body problem).
	 We assume $\Delta=\delta$ to be real throughout all derivations. Then, we analytically continue the result to the complex case $\Delta = \delta - i\gamma$. Throughout all derivations, we also consider the situation where the photon field is on the EIT resonance ($\delta_r=0$) and the Rydberg state  decay rate is negligible ($\gamma_s = 0$).
	 
    \par Spectral decomposition of the spin-wave part of the Hamiltonian \eqref{Seq:H1M} gives energies $\epsilon_{\pm}$ of the upper/lower spin wave and their respective overlaps $\alpha_{\pm}$ with the $\ket{s}$ state:
    \begin{align}
        \label{Seq:Ealpha}
        \epsilon_{\pm} &= \frac{1}{2} \left(\Delta\pm\sqrt{\Delta ^2+4\, \Omega^2}\right), & \alpha_{\pm} &= \frac{1}{2}\left(1\pm\frac{\Delta }{  \sqrt{\Delta ^2+4\,\Omega ^2}}\right).
    \end{align}
    In the coordinate space, the single-particle propagator ($ \hat{g}_s $ in the main text) is 
    \begin{equation}\label{Seq:gs}
    	 g_s(\omega;x,x')=\underbrace{\sum_{\mu=\pm} \frac{\alpha_\mu}{\omega-\epsilon_{\mu}+i0^+}}_{\eta(\omega)} \sum_{q\neq 0}e^{iq(x-x')}+\frac{1}{\omega+i0^+},
    \end{equation}
    where the first term corresponds to the spin-wave excitation and the second to the dark-state polariton.   In the large-coupling limit $g\gg\Omega,\abs{\Delta}$, only the dark branch of polaritons contributes because the energy of bright polaritons is proportional to $g \rightarrow \infty$, so their effect %might
    can be neglected -- see Ref.~\cite{Bienias2014} for a more detailed discussion. The energy-dependent factor $ \eta(\omega) $ can be evaluated to
      \begin{equation}
    \eta(\omega) = \frac{\omega-\Delta }{\omega^2-\left(\Omega^2+\Delta\,  \omega\right)},
    \end{equation}
    and one needs to use the full expression with $ i0^+ $ included if integration occurs.
    
    \par In the previous work of Refs.~\cite{Jachymski2016,Gullans2016}, the authors considered the limit  $\Omega\ll\Delta$, where only one spin-wave branch contributes.   However, in our regime of interest $\Omega\sim\Delta$ and we must solve the more general problem that includes all branches. In addition, both Ref.~\cite{Jachymski2016} and Ref.~\cite{Gullans2016} made (different) simplifying approximations which led to slight quantitative differences between all three  solutions for the effective three-body force: Ref.~\cite{Jachymski2016}, Ref.~\cite{Gullans2016}, and the present work. The current approach is more systematic and can be rigorously derived as an asymptotic perturbative expansion of the solutions to the three-body Schr{\"o}dinger equation.  We present further extensions of these results to general multi-mode cavities in our upcoming work \cite{Kalinowski2020}.  Generalizing these cavity solutions to the free-space problem remains an outstanding challenge. An alternative technique, that has been successfully applied in free-space, is to define effective three-body parameters through nonperturbative matching techniques \cite{Liang2018}.  In this alternative approach, these parameters are tuned in an effective field theory to reproduce low-energy observables (e.g., the dimer-polariton scattering length) obtained from the solution to the microscopic model.
    
    \section{Two-body Problem}\label{Ssec:2}
    In this section, we introduce notation important for our scattering analysis and solve the two-body problem.
    
 \par To gain insight into the influence of the additional spin-wave branch, we begin by studying the two-body problem. Description of the two-body processes is contained in the off-shell two-body T-matrix $T_2(\omega)$. Here, we derive Eq.~\meq{eq:T2eq}{3} from the main text. An equation for $T_2(\omega)$ can be written explicitly using a supporting definition $T_2(\omega)=\int \frac{dr}{L}T_2(\omega;r)$, where $T_2(\omega;r)=\int \frac{d \vec{x}'}{L^2} \frac{dR }{L} T_2(\omega;\vec{x},\vec{x}')$, $ \vec{x}=(x_1,x_2) $, $ \vec{x}'=(x_1',x_2') $, $ r = x_1 - x_2$ is the relative coordinate, and $R = (x_1 + x_2)/2$ is the center of mass coordinate. $T_2(\omega;r)$ is given by
    \begin{equation}
        \label{Seq:T2LS}
        T_2(\omega;r) = V(r) +\int \frac{dr'}{L}\, V(r) G_2(\omega;r,r') T_2(\omega;r'),
    \end{equation}
    where the two-body propagator $G_2$ can be obtained from the Hamiltonian \eqref{Seq:H1M} and is given by
    \begin{equation}\label{Seq:G2ini}
        G_2(\omega;r,r') = (\omega+i0^+)^{-1} + \chi(\omega)\sum_{q\neq0} e^{iq (r-r')},
    \end{equation}
    where the zero-momentum term $(\omega+i0^+)^{-1}$ is the propagator of two dark-state polaritons. It can be related to our formulation using $ \hat{g}_s $ via
    \begin{equation}\label{Seq:G2fromgs}
     G_2(\omega;r,r')\overset{!}{=}\int \frac{d\omega'}{2\pi i}g_s(\omega';x_1,x_1')g_s(\omega-\omega';x_2,x_2'),
    \end{equation}
    where $ \overset{!}{=} $ means that it is equal only under the integral in Eq.~\eqref{Seq:T2LS} and after enforcing the total-momentum conservation.
    The coefficient $\chi(\omega)$, in the part of $G_2$ corresponding to the double excitation of spin waves, is
    \begin{equation}
        \chi(\omega) = \frac{\Delta - \frac{\omega}{2}-\frac{\Omega^2}{\Delta-\omega}}{\omega\left(\Delta-\frac{\omega}{2}\right)+2\Omega^2},
    \end{equation}
    which coincides with the two-body propagator in free space in the infinite-momentum limit \cite{Bienias2014}.
    Equation \eqref{Seq:T2LS} is represented schematically in Fig.~\mref{fig:T2T3diag}{2}(a) of the main text. To solve it, we rewrite the propagator from Eq.~\eqref {Seq:G2ini} as
    \begin{equation}
        G_2(r,r') = [(\omega+i0^+)^{-1}-\chi(\omega)]+L\chi(\omega)\, \delta(r-r')
    \end{equation}
    and plug it back into Eq.~\eqref{Seq:T2LS}. After rearrangement and integration of both sides by $\int \frac{dr}{L}$, we obtain the equation for $T_2(\omega)$, which gives
    \begin{equation}\label{Seq:T2eff}
        T_2(\omega) = U_2(\omega) + U_2(\omega)[(\omega+i0^+)^{-1}-\chi(\omega)]T_2(\omega),
    \end{equation}
    where $U_2(\omega)$ is a well-known \cite{Bienias2014} renormalized two-body interaction (for $g\rightarrow \infty$) between dark-state polaritons, presented in Eq.~\meq{eq:Veff2}{2} of the main text. In the limit of large separation, it reduces to the bare van der Waals potential. Conversely, for small distances, it saturates at a finite value, an effect caused by the so-called Rydberg blockade mechanism.
    
    \par Solving algebraic equation \eqref{Seq:T2eff} for $T_2(\omega)$ reproduces Eq.~\meq{eq:T2eq}{3} from the main text. Notice that, to the leading order in $r_b/L$, equation \eqref{Seq:T2eff} describes the scattering of two infinitely heavy particles under the influence of potential $U_2(\omega;r)$. Additional terms encapsulate effects specific to the cavity setup. 
    
    \section{Three-body Problem}\label{Ssec:3}
    In this section, we investigate the scattering of three dark-state polaritons relying heavily on definitions and results from the previous two sections. Schematic representation of the key equations is presented in Fig.~\mref{fig:T2T3diag}{2}(b-c) of the main text.

    \par As in the two-body case, the influence of the interaction on the physics of the three-body problem is captured by the analytical structure of the T-matrix $\bm{T_3}$. Specifically, the energy shift corresponds to the pole of the integrated T-matrix $T_3(\omega)$. 
    
    \par To calculate this object, one can in principle employ the Schr\"{o}dinger equation. However, such a treatment can lead to spurious, nonphysical solutions. In order to avoid this issue, another approach to the quantum three-body problem was developed by Faddeev  \cite{Faddeev1961}. In this formulation, all scattering processes are grouped depending on which two particles interact first. This way one introduces a rigorous method for expressing the three-body scattering as a series of two-body processes (a three-body force can also be included). As a consequence, the three-body T-matrix can now be written as a sum of  three (sub)T-matrices
    \begin{equation}
    \hat{T}_3(\omega)=\frac{1}{3}\sum_{i<j}\hat{T}_3^{ij}(\omega,\epsilon_k),
    \end{equation}
    where $ \hat{T}_3^{ij}(\omega;\epsilon_k) $ denotes the T-matrix for the group of processes, where particles labeled $ i $ and $ j $ interacted first and the third particle $ k\neq i,j $ has energy $ \epsilon_k $. These $ \hat{T}_3^{ij}(\omega,\epsilon_k) $  objects are coupled to each other by the set of equations called Faddeev equations. In our case, the situation is further complicated by the multi-component nature of the polariton system. Let us define $ \hat{T}_3^{ij}(\omega)\equiv \hat{T}_3^{ij}(\omega;0) $ and $ \hat{T}_{\mu}^{ij}(\omega)\equiv \hat{T}_3^{ij}(\omega;\epsilon_{\mu=\pm}) $.
    
   \par Every T-matrix we consider has all three outgoing DSPs. In our system, exact Faddeev equations \cite{Faddeev1961} describing the off-shell scattering  of three zero-momentum DSPs are
   \begin{align}
   T_3^{12}(\omega) &= T_2(\omega) \frac{2}{\omega+i0^+}T_2(\omega)+T_2(\omega)\frac{1}{\omega+i0^+}\left(T_3^{13}(\omega)+T_3^{23}(\omega)\right)\nn\\
   &+\sum_{\mu=\pm} \int \frac{dr_{12}}{L}\frac{dr_3}{L} T_2(\omega;r_{12})\,\alpha_\mu \eta_\mu  \left(T_\mu^{13}(\omega;\vec{x})+T_\mu^{23}(\omega;\vec{x})\right)-T_2(\omega)\,\alpha_\mu \eta_\mu  \left(T_\mu^{13}(\omega)+T_\mu^{23}(\omega)\right)\label{Seq:T3exact}
   \end{align}
   and the two analogous equations for $ T_3^{13}(\omega)$ and $ T_3^{23}(\omega)$ obtained through permutation of indices. Here $T^{ij}_{\mu}(\omega;\vec{x}) = \int \frac{d\vec{x}'}{L^3}\frac{dR}{L}T^{ij}_\mu(\omega;\vec{x},\vec{x}')$ is the T-matrix for the process with the third incoming leg being a spin wave belonging to the branch $\mu$. Variable $r_{12}$ denotes the distance between particles labeled 1 and 2, $r_3=(x_1+x_2)/2-x_3$ is the standard third Jacobi coordinate, and $\eta_\mu = \eta(\omega-\epsilon_\mu)$.
   
      \par Eq.~\eqref{Seq:T3exact} is a direct implementation of the operator equation [Eq.~\meq{eq:Fadeev}{6}] from the main text. Here, we explicitly separated spin-wave and DSP terms of $ \hat{g}_s $  as in Eq.~\eqref{Seq:gs} and then performed the $ \tilde{\epsilon} $ integration with the enforcement of momentum invariance. Note that, in contrast to the two-body problem, here we do not obtain an effective few-body propagator as $ \chi(\omega) $, since the three-body T-matrix depends on the energy of the third particle and prevents such grouping.
   
   \par Now we introduce our main approximation, which allows us to write a self-consistent system of equations to the lowest non-trivial order in $r_b/L$. For this purpose, we keep only those terms where either the sum over a macroscopic number of spin waves is present or the all-dark intermediate state arises. The sum over a macroscopic number of spin waves introduces a factor of $L\,\delta(x)$, while the all-dark-state propagator contributes $(\omega+i0^+)^{-1}$, which is of order $L/r_b$. In comparison, all other terms are negligible in the limit of vanishing  $r_b/L$, as they are of order $ \sim 1  $ or smaller. The key consequence of these rules is as follows: $ \{T_3\} =1 $ and $ \{T_\mu \}=2$, where $\{X\}$ denotes the leading order of  $X$ in $ r_b/L $

   \par For clarity, we will omit terms that arise from  higher-order corrections to the  $ g_{s} $ propagator (such as the last term in Eq.~\eqref{Seq:T3exact}) as they do not contribute to our lowest-order solution. We note that, in general, the $ T_2 $ matrix depends on the total momentum of two particles $ K $, but in all our equations we can either truncate it to $ U_2 $, or it arises in a situation (outer $ T_2 $ contributions) where $ K=0 $.
   
   \par In order to close the system of equations, we need to calculate $T_\mu^{ij}$, which is described by another set of Faddeev equations expanded to leading order in $r_b/L$ 
   \begin{align}
   T_\mu^{12}(\omega,\vec{x}) =& \sum_{b\neq(12)(3)} \left[U^\mu_2(x_1-x_2)\eta_\mu\,T_2(\omega;x_{b_1}-x_{b_2})+\sum_{\nu=\pm}\alpha_\nu\,U_2^\mu(x_1-x_2)\eta_{\mu\nu}T^{b_1b_2}_\nu(\omega;\vec{x})\right]\nn\\
   &+\sum_{b\neq(12)(3)}\,U^\mu_2(x_1-x_2)\eta_\mu\,T_2(\omega;x_{b_1}-x_{b_2})\frac{1}{\omega+i0^+}\left[2\,T_2(\omega)+\sum_{c\neq(b_1 b_2)(b_3)}T_3^{c_1 c_2}(\omega)\right],\label{Seq:Tmufull}
   \end{align}	
   where $U_2^\mu(x)\equiv U_2(\omega-\epsilon_{\mu};x)$ and $\eta_{\mu\nu} = \eta(\omega-\epsilon_\mu-\epsilon_\nu)$. It is easier to first calculate $\tilde{T}_\mu^{12}$, where the tilde denotes that it is a T-matrix where the intermediate DSPs come only from nonperturbative corrections to $ T_2 $. This is given by
   \begin{equation}
   \tilde{T}_\mu^{12}(\omega,\vec{x}) = \sum_{b\neq(12)(3)} \left[U^\mu_2(x_1-x_2)\eta_\mu\,T_2(x_{b_1}-x_{b_2})+\sum_{\nu=\pm}\alpha_\nu\,U_2^\mu(x_1-x_2)\eta_{\mu\nu}\tilde{T}^{b_1b_2}_\nu(\omega;\vec{x})\right]
   \label{Seq:Tmutil}
   \end{equation}
   and is related to Eq.~\eqref{Seq:Tmufull} by 
   \begin{equation}
   T_\mu^{12}(\omega;\vec{x}) = \tilde{T}_\mu^{12}(\omega;\vec{x})+ \tilde{T}_\mu^{12}(\omega;\vec{x})\frac{1}{\omega+i0^+}\left(T_3^{13}(\omega)+T_3^{23}(\omega)+2 \,T_2(\omega)\right).  \label{Seq:Tmuldsp}
   \end{equation}
 This solution can be verified by inserting Eqs.~\eqref{Seq:Tmutil}~and~\eqref{Seq:Tmuldsp} into Eq.~\eqref{Seq:Tmufull} and seeing that everything cancels out (we also use $ T_3^{ij}(\omega)=T_3^{kl}(\omega) $ for any $ i\neq j $, $ k\neq l $).
The system of algebraic equation in Eq.~\eqref{Seq:Tmutil} can be solved analytically.
   Inserting this result into Eq.~\eqref{Seq:T3exact} and summing over all pairs $i,j$ gives the renormalized system of equations for the dark-state T-matrix
   \begin{equation}\label{Seq:T3full}
   T_3(\omega) =  T_2(\omega) \frac{2}{\omega+i0^+}T_2(\omega)+T_2(\omega)\frac{2}{\omega+i0^+}T_3(\omega)+ \Phi^2\, U_3(\omega) +\Phi^2\, U_3(\omega)\frac{2}{\omega+i0^+}\left[T_3(\omega)+T_2(\omega)\right] + \mathcal{O}(r_b^3/L^3),
   \end{equation}
   where $ \Phi = T_2(\omega)/U_2(\omega) $ is the non-perturbative correction. The effective three-body potential $U_3(\omega)$ can be concisely written as
   \begin{equation}
   U_{\rm 3}(\omega) = \int \frac{dx}{L}\frac{dy}{L}\, \sum_{\mu=\pm} 2\,  U_2(\omega;x-y) \,\eta_\mu \alpha_\mu\,\widehat{T}_\mu^{12}(\omega;x,y),
   \end{equation}
   which was possible due to symmetries with respect to the relabeling of particles and coordinates. Variables $x=x_1-x_2$ and  $y=x_3-x_2$ describe relative distances of pairs of particles and $ \widehat{T}_\mu^{12}(\omega;x,y)= \tilde{T}_\mu^{12}(\omega;x,y)/\Phi$ is governed by Eq.~\eqref{Seq:Tmutil} with $ T_2\rightarrow U_2 $. 
   The function  $U_3(\omega)$  can be intuitively understood as an effective three-body potential, with direct analogy to $U_2(\omega)$ in the two-body scenario.
   
\par Solving Eq.~\eqref{Seq:T3full} for $T_3(\omega)$, we get 
\begin{equation}\label{Seq:T3eq}
T_3(\omega) = \frac{2 T_2(\omega)^2+(\omega+2T_2(\omega)) \,U_{3}/(1-U_2[1-\chi(\omega)\,\omega])^2}{\omega-2\left(T_2(\omega)+U_{3}(\omega)/(1-U_2(\omega)[1-\chi(\omega)\,\omega])^2\right)},
\end{equation}
Its pole gives the equation for the three-body energy shift: 
\begin{equation}
    \delta E_3 =3U_2(\delta E_3) + 3U_{3}(\delta E_3) - \chi(0) U_2(\delta E_3)\,\delta E_3+\mathcal{O}(r_b^3/L^3)\label{Seq:E3eq}.
\end{equation}
To obtain a self-consistent solution, we must expand both sides to the appropriate order in $r_b/L$. For this purpose, we analyze the order of each constituent: 
\begin{align}
\{\delta E_3\}&=1,&\{U_2\}&=1,&\{U_3\}&=2,&\{\chi(0)\}&=0.
\end{align}
 This allows us to write the final solution
\begin{equation}
\delta E_3 = \underbrace{3U_2}_{\mathcal{O}(r_b/L)}+\underbrace{3U_3+3U_2\left(3U_2'-\chi\,U_2\right)}_{\mathcal{O}(r_b^2/L^2)} +\mathcal{O}(r_b^3/L^3),
\end{equation}
where $U_2' = d U_2/d\omega$ and all functions are evaluated at $\omega=0$. Notice  that the multi-branch character of the problem is contained in  $U_3$ and in the more complicated form of $\chi$ compared to the regime  $\Omega\ll\abs{\delta}$.

    \section{Transmission calculations}\label{Ssec:4}
    In this section, we present the model describing the transmission of photons in a chiral waveguide coupled to a single-mode cavity described by Eq.~\meq{eq:Htr}{7} in the main text. We analytically calculate the three-body loss parameter $r_3$ in the case when $u_2=\kappa=0$, which corresponds to the situation where three-body effects dominate. The results presented in this section are used to obtain Fig.~\mref{fig:transmission}{4}(a-b) in the main text.
    
   \par The whole system is described by the transmission Hamiltonian %$H_{\rm tr}$, 
    \begin{align*}
    H_{\rm tr}=\int_{-\infty}^{+\infty} dk\, k\, C^\dagger(k) C(k) +\int_{-\infty}^{+\infty} dk\,\sqrt{2\pi} g\left(b^\dagger C(k)+C^\dagger(k)b\right)+u_3\, (b^\dagger)^3 b^3,
    \end{align*}
   where the first term describes the Hamiltonian of the photons in the chiral waveguide. $C^\dagger(k)$ and $C(k)$ are creation and annihilation operators of chiral photons at momentum $k$, respectively. Speed of light is set to unity ($c=1$). The second term describes the quadratic coupling between photons in the waveguide and the cavity, where $b^\dagger$ ($b$) creates (destroys) a cavity photon. The last term describes the three-body nonlinear interactions of  cavity photons.

   \par To compute few-photon scattering, it is convenient to partition the Hamiltonian into the quadrtic part $H_0$ and the nonlinear interactions:
     \begin{align*}
    H_{\rm tr}&\equiv H_0+U, \\
    H_0&=\int_{-\infty}^{+\infty} dk\, k\, C^\dagger(k) C(k) +\int_{-\infty}^{+\infty} dk\,\sqrt{2\pi} g\left(b^\dagger C(k)+C^\dagger(k)b\right),\\
    U&=u_3\, (b^\dagger)^3(b)^3. 
    \end{align*}

    \par The quadratic part $H_0$ of the Hamiltonian $H_{\rm tr}$ can be  diagonalized into the scattering eigenstates:
      \begin{align*}
      H_0 &= \int_{-\infty}^{+\infty} dk \,k \psi_k^\dagger \psi_k,\\
     \psi^\dagger_k &= e_k b^\dagger + \int_{-\infty}^{+\infty} dk'\,\psi_k(k')C^\dagger(k),
    \end{align*}
    where $ e_k = \frac{1}{\sqrt{2\pi}}\frac{g}{k+i\kappa +i \Gamma}$ and $[\psi_k,\psi_{k'}^\dagger]=\delta(k-k')$. $\{\psi_k^\dagger|k\in ( {-\infty,+\infty})\}$ form a complete basis of the Hilbert space, which we refer to as the dressed-photon basis. 
    Let $\psi_k(z)$ be the Fourier transform of $\psi_k(k')$ in the coordinate space. The asymptotic behavior of $\psi_k(z)$ gives the single-photon transmission coefficient 
    \begin{equation}
    t_k = \frac{\lim_{{z\rightarrow }+\infty} \psi_k(z)}{\lim_{{z\rightarrow }-\infty} \psi_k(z)}=\frac{k-i\Gamma}{k+i\Gamma}. \label{eqtk}
    \end{equation}
    
 \par The nonlinear interaction $U$ can also be expressed in the dressed photon basis: 
 \begin{align}
 U&=\int d\vec{k} d\vec{k'}\ U(\vec{k},\vec{k'}) \psi^\dagger_{k'_1}\psi^\dagger_{k'_2}\psi^\dagger_{k'_3}\psi_{k_1}\psi_{k_2}\psi_{k_3},\label{eqU3}\\
 U(\vec{k},\vec{k'})&=e^*_{\vec{k'}}e_{\vec{k}}u_3,
\end{align}
where we have used the definitions $\vec{k}=(k_1,k_2,k_3)$ and 
$e_{\vec{k}}=e_{k_1}e_{k_2}e_{k_3}$. 
Note that, in the dressed-photon basis, $ U(\vec{k},\vec{k'})$ is expressed as the product of separable functions of the incoming and outgoing momenta. We will use this feature to introduce a simple ansatz for the three-body dressed-photon T-matrix, which is a solution to the Lippmann-Schwinger equation:
\begin{equation}
T^{(3)}(\omega, \vec{k},\vec{k'})=U(\vec{k},\vec{k'})+\int_{-\infty}^{+\infty} d\vec{k''} \frac{U(\vec{k},\vec{k''})}{\omega-k}T^{(3)}(\omega,\vec{k}, \vec{k''}).\label{eqLST3}
\end{equation}
Proposing an ansatz $T^{(3)}(\omega, \vec{k},\vec{k'})=\bar{T}(\omega)e^*_{\vec{k'}}e_{\vec{k}}$ and inserting it into Eq.~\eqref{eqLST3}, we find 
\begin{equation}
    	\bar{T}(\omega)=\frac{1}{\frac{1}{u_3}-\frac{1}{\omega+3i\Gamma}},
\end{equation}
where we have assumed that Im$[\omega]>0$.

\par Hence we obtain the three-photon S-matrix for the dressed photons:
 \begin{equation}
S^{(3)}( \vec{k},\vec{k'}) = \delta(\vec{k}-\vec{k'})-2\pi i\delta(K-K') T^{(3)}(K+i0, \vec{k}, \vec{k'}) \label{eqS3d},
    \end{equation}
which represents the transmission amplitude of incoming dressed photons with energy $\vec{k}=(k_1,k_2,k_3)$ scattered into outgoing dressed photons with energy  $\vec{k'}=(k'_1,k'_2,k'_3)$. $K$ and $K'$ are the total momenta/energies of the incoming and outgoing photons: $K=k_1+k_2+k_3$, $K'=k'_1+k'_2+k'_3$. Note that the S-matrix $S_0^{(3)}$ for free photons (as opposed to dressed photons) is more relevant for direct experimental measurements and can be obtained from the dressed-photon S-matrix $S^{(3)}$ using single-photon transmission coefficients: $
     S_0^{(3)}(\vec{k},\vec{k'})  = t_{k_1'} t_{k_2'} t_{k_3'} S^{(3)}( \vec{k},\vec{k'})$.

\par Next, we calculate $r_3$ defined in Eq.~\meq{eq:r3}{8} in the main text and representing a good measure of three-body loss in the case of zero one- and two-body losses:

\begin{equation}
r_3=\int_{-\infty}^{+\infty} d\tau_1 d\tau_2\,[ 1-g^{(3)}(\tau_1,\tau_2)].
\label{Seq:r3}
\end{equation}
$g^{(3)}(\tau_1,\tau_2)$ is the three-photon correlation function at the output of the waveguide, where $\tau_{1}=t_2-t_3, \tau_2=t_1-t_3$ are the  time differences between the photon number measurements at times $t_{1,2,3}$. For a weak and continuous coherent-state input with photon momentum $k$, $g^{(3)}(\tau_1,\tau_2)$ is related to the output three-photon wavefunction $\psi^{(3)}(z_1=t_1,z_2=t_2,z_3=t_3)$ in the dressed-photon basis, which is the Fourier transform of $S^{(3)}( \vec{k}=(k,k,k), \vec{k}')$ with respect to the output momenta $\vec{k'}$.

  \par Defining $R = (z_1 + z_2 + z_3)/3$ as the center of mass coordinate, we have
  \begin{align}
  \psi^{(3)}(z_1,z_2,R)=\exp(3ikR) (1-\phi^{(3)}(z_1,z_2)),\label{eqpsi3phi3}
  \end{align}
where $\phi^{(3)}(z_1,z_2)$ is the Fourier transform of $y(k'_1, k'_2)=2\pi i T^{(3)}(3k+i0, \vec{k},k'_1, k'_2, k'_3=3k-k'_1-k'_2)$ with respect to $k'_1, k'_2$. $g^{(3)}(\tau_1,\tau_2)$ is then given by 
\begin{equation}
	g^{(3)}(\tau_1,\tau_2)=|\psi^{(3)}(z_1,z_2, R)|^2=|1-\phi^{(3)}(z_1,z_2)|^2, \label{eqg3psi3}
\end{equation}
where $z_1=-\frac{1}{3}\tau_1+\frac{2}{3}\tau_2$, $z_2=\frac{2}{3}\tau_1-\frac{1}{3}\tau_2$. Hence $r_3$ defined in Eq.~\eqref{Seq:r3} can be calculated as follows:
  \begin{align}
  r_3&= \int_{-\infty}^{+\infty} dz_1 dz_2\  [1- |1-\phi^{(3)}(z_1,z_2)|^2]\\
  &= \int_{-\infty}^{+\infty}dz_1 dz_2 \ 2\text{Re}[\phi^{(3)}(z_1,z_2) ]-| \phi^{(3)}(z_1,z_2)|^2 \label{eqline3}\\
  &=2\text{Re}[ y(k'_1=k, k'_2=k)]-\int_{-\infty}^{+\infty} dk'_1dk'_2 |y(k'_1, k'_2)|^2\label{eqline4}\\
  &=4\pi \left(\frac{\Gamma}{\pi}\frac{1}{k^2+\Gamma^2}\right)^3\frac{\text{Im}[-\frac{1}{u_3^*}]}{\left\lvert\frac{1}{u_3}-\frac{1}{3k+3i\Gamma}
\right\rvert^2}\geq 0.\label{eqline5}
  \end{align}

\par We comment that the calculation of $r_3$ presented here for $\kappa=u_2=0$ is a special case of a general calculation applicable to arbitrary values of $\kappa,u_2$  and presented in an upcoming work on cavity transmission \cite{Wang2020}.

%</document>
    \bibliography{library.bib}